\scrollmode
\documentstyle[12pt]{article}
\newif\ifdraft 
 \drafttrue
 \draftfalse
\newif\ifappendices \appendicesfalse

\begin{document}
\pretolerance=5000
\tolerance=5000
\parskip=0pt plus 1pt
\parindent=24pt
%\pageno=1
\tolerance=10000
\hfuzz=5pt

\def\thebibliography#1{{\bf\noindent References}\vspace{2.3ex plus .2ex}
 \list {[\arabic{enumi}]}{\settowidth\labelwidth{[#1]}\leftmargin\labelwidth
 \advance\leftmargin\labelsep
 \usecounter{enumi}}
 \def\newblock{\hskip .11em plus .33em minus .07em}
 \sloppy\clubpenalty4000\widowpenalty4000
 \sfcode`\.=1000\relax}
\let\endthebibliography=\endlist
%
% Heading for list of references
%
\def\references
     {\vskip-\lastskip
    \vskip36pt plus12pt minus12pt
    \bigbreak
    \vfill\eject
     {\noindent \bf References\par}
      \parindent=0pt
      \bigskip}
\def\figures
   {\vskip-\lastskip
    \vskip36pt plus12pt minus12pt
    \bigbreak
    \noindent{\bf Figure Captions \par}
    \nobreak
    \bigskip
    \noindent}
% Heading for acknowledgments
%
\def\ack
   {\vskip-\lastskip
    \vskip36pt plus12pt minus12pt
    \bigbreak
    \noindent{\bf Acknowledgments\par}
    \nobreak
    \bigskip
    \noindent}
%

%Define or redefine some commands for title page:
\def\fncomma{$^{\rm ,}$}
\renewcommand{\footnoterule}{}
\renewcommand{\footnotesep}{12pt}
\let\thanks\footnote
\catcode`\@=11
\def\@makefntext#1{\setbox0\hbox{$^{\@thefnmark}\,$}%
  \hangindent\wd0\hangafter1\noindent\box0\ignorespaces#1}
\catcode`\@=12
%
% counter definitions
%
%\newcount\secno      %section number
\newcount\subno      %number of subsection
\newcount\subsubno   %number of subsubsection
\newcount\appno      %appendix number
\newcount\appeqno    %appendix number
\newcount\tableno    %table number
\newcount\figureno   %figure number
\newtheorem{theorem}{Theorem}[section]
\newtheorem{lemma}{Lemma}

%
% Section heading (#1 is title of section, no number required)
%
\def\section#1
   {\vskip0pt plus.1\vsize\penalty-250
    \vskip0pt plus-.1\vsize\vskip24pt plus12pt minus6pt
    \subno=0 \subsubno=0
    \setcounter{equation}{0}
    \addtocounter{section}{1}
%    \global\advance\secno by 1
    \noindent {\bf \thesection. #1\par}
    \bigskip
    \noindent}
\def\appendix#1
   {\vskip0pt plus.1\vsize\penalty-250
    \vskip0pt plus-.1\vsize\vskip24pt plus12pt minus6pt
    \subno=0 \subsubno=0
    \setcounter{equation}{0}
    \ifappendices\else\setcounter{section}{0}\fi\appendicestrue
    \addtocounter{section}{1}
    \def\thesection{\Alph{section}}
    \noindent {\bf Appendix \thesection. #1\par}
    \bigskip
    \noindent}
\renewcommand{\theequation}{\thesection.\arabic{equation}}
%\renewcommand{\theequation}{section}{\arabic{equation}}
%\def\theequation{\arabic{secno}.\arabic{eqno}}
%
% Subsection heading (#1 is title of subsection, no number required)
%
\def\subsection#1
   {\vskip-\lastskip
    \vskip24pt plus12pt minus6pt
    \bigbreak
    \global\advance\subno by 1
    \subsubno=0
    \noindent {\sl \thesection.\the\subno. #1\par}
    \nobreak
    \medskip
    \noindent}
\def\subsection*#1
   {\vskip0pt plus.1\vsize\penalty-250
    \vskip0pt plus-.1\vsize\vskip24pt plus12pt minus6pt
    \subno=0
    \noindent {\bf  #1\par}
    \noindent}

% Subsubsection heading (#1 is title of subsubsection,
% no number is required)
%
\def\subsubsection#1
   {\vskip-\lastskip
    \vskip20pt plus6pt minus6pt
    \bigbreak
    \global\advance\subsubno by 1
    \noindent {\sl \thesection.\the\subno.\the\subsubno. #1}\null. }

%% DEFINITION OF CONTROL SEQUENCES FOR MATHEMATICS
\def\qed{\nobreak\kern 1em \vrule height .5em width .5em depth 0em}
\def\vbar{\mathchoice{\vrule height6.3ptdepth-.5ptwidth.8pt\kern-.8pt}
   {\vrule height6.3ptdepth-.5ptwidth.8pt\kern-.8pt}
   {\vrule height4.1ptdepth-.35ptwidth.6pt\kern-.6pt}
   {\vrule height3.1ptdepth-.25ptwidth.5pt\kern-.5pt}}
\def\bbr{{\rm I\mkern-3.5mu R}}
\def\bbc{{\rm \mkern5mu\vbar\mkern-5mu C}}
\def\bbz{{\rm Z\mkern-8mu Z}}
\def\Re{\mathop{\rm I\mkern-3mu R}}
\def\(#1){(\ref{#1})}
\def\[#1]{\cite{#1}}
\def\E#1{\langle#1\rangle}
\def\EE#1{\left\langle#1\right\rangle}
\def\Pr{\mathop{\rm Prob}}
\def\px{\partial_x}\def\pt{\partial_t}
\def\a{\sqrt a}\def\b{\sqrt b}
\def\v{|v\rangle}\def\w{\langle w|}
\def\Ewv#1{\langle w|#1|v\rangle}
\def\sech{\mathop{\rm sech}\nolimits}
\def\sp{\mathop{\displaystyle\sum\nolimits^{\,\prime}}}
\def\spp{\mathop{\displaystyle\sum\nolimits^{\,\prime\prime}}}
\def\d{\hat D}\def\e{\hat E}
\def\xxx{\mathrel{\mathop{\longrightarrow}\limits_{\epsilon\to0}}}

%% END OF CONTROL SEQUENCES FOR SYMBOLS

 \def\title{Shock Profiles for the Asymmetric Simple Exclusion Process 
     in One Dimension}

 \def\author{B.\ Derrida\thanks{\small Laboratoire de Physique Statistique,
Ecole Normale Sup\'erieure, 24 rue Lhomond, 75005 Paris, France; 
email derrida@lps.ens.fr.},
 J.\ L.\ Lebowitz\thanks{\small Institut des Hautes Etudes Scientifiques,
Le Bois-Marie, 35 route de Chartres, F--91440 Bures-sur-Yvette, France.
Permanent address: Department of Mathematics,
Rutgers University, New Brunswick, NJ 08903; email lebowitz@math.rutgers.edu,
speer@math.rutgers.edu.},
 E.\ R.\ Speer$^{\tiny 2}$}

 \def\abstracttext{The asymmetric simple exclusion process (ASEP) on a
one-dimensional lattice is a system of particles which jump at rates
$p$ and $1-p$ (here $p>1/2$) to adjacent empty sites on their right
and left respectively.  The system is described on suitable
macroscopic spatial and temporal scales by the inviscid Burgers'
equation; the latter has shock solutions with a discontinuous jump
from left density $\rho_-$ to right density $\rho_+$, $\rho_-<\rho_+$,
which travel with velocity $(2p-1)(1-\rho_+-\rho_-)$.  In the
microscopic system we may track the shock position by introducing a
second class particle, which is attracted to and travels with the
shock.  In this paper we obtain the time invariant measure for this
shock solution in the ASEP, as seen from such a particle.  The mean
density at lattice site $n$, measured from this particle, approaches
$\rho_{\pm}$ at an exponential rate as $n\to\pm\infty$, with a
characteristic length which becomes independent of $p$ when
$p/(1-p)>\sqrt{\rho_+(1-\rho_-)/\rho_-(1-\rho_+)}$.  For a special
value of the asymmetry, given by
$p/(1-p)=\rho_+(1-\rho_-)/\rho_-(1-\rho_+)$, the measure is Bernoulli,
with density $\rho_-$ on the left and $\rho_+$ on the right.  In the
weakly asymmetric limit, $2p-1\to0$, the microscopic width of the
shock diverges as $(2p-1)^{-1}$.  The stationary measure is then
essentially a superposition of Bernoulli measures, corresponding to a
convolution of a density profile described by the viscous Burgers
equation with a well-defined distribution for the location of the
second class particle.}

\def\date{\today}

\begin{titlepage}

\begin{center}{\LARGE \title\par}

\vskip1em
{ \large \author\par}

\ifdraft \vskip1em
DRAFT \date\par\fi
 \end{center}
 \vskip2em

 \vfill
 \begin{abstract}
 \ifdraft ABSTRACT MOVED TO SECOND PAGE IN DRAFT VERSION\else\abstracttext\fi
 \end{abstract}

 \vfill
 \vskip2em
 \noindent
% {\bf Date:} 4/21/96

 \vskip0.5em
 \noindent 
 {\bf Submitted to:} {\it Journal of Statistical Physics}


\vskip1em
\eject

%%TEMPORARY PAGE TO PERMIT GENEROUS SPACING OF ABSTRACT IN DRAFT VERSION
 \ifdraft\count0=0
 {\def\folio{}
 \openup5\jot
 \begin{abstract} 
 \abstracttext
 \end{abstract}
 \openup-5\jot
 \vfill\eject}\fi

\end{titlepage}

\baselineskip=18pt plus 3pt minus 3pt
\ifdraft\openup3\jot\fi

\section{Introduction}

The aim of this and our previous work~\[djls,DJLS], with S.~Janowsky,
is the determination of the underlying microscopic structure of a
fluid in regions in which it has shocks on the macroscopic scale.  We
consider situations in which the system evolves macroscopically
according to some deterministic autonomous equations, such as the
Euler, Navier-Stokes, or Burgers' equations~\[Spohn,MP,LPS].  When the
solutions of these equations are smooth, we can assume that on the
microscopic level the system is (essentially) in a local equilibrium
state, determined by the local macroscopic parameters obtained from
the solutions.  What is less clear, however, and is of particular
interest, is what happens when the macroscopic evolution is not
smooth, as in the occurrence of shocks; these are described by regions
of very large gradient in solutions of the Navier-Stokes or viscous
Burgers' equations and by discontinuities in solutions of their zero
viscosity limits, the Euler or inviscid Burgers' equations.  It is
first of all not clear in what sense the macroscopic equations are to
be interpreted in such regions, since their derivation (heuristic or
rigorous) is based on the assumption of slow variation in the system's
properties on the microscopic scale.  Beyond that, these equations do
not describe the structure of the shocks on the microscopic scale.
This is the problem we wish to address here.  In particular, do the
statistical properties of the atoms or molecules change abruptly on
the interparticle distance scale? Or do these properties change
significantly only over much larger distances? (For a discussion of
shock structure in gases based on the ``mesoscopic'' description
provided by the Boltzmann equations, see \[Cer].)

One model for which these questions have been answered, at least partially,
is the asymmetric simple exclusion process (ASEP)~\[Spi,Ligg] on the
one-dimensional lattice $\bbz$.  In this model particles attempt at random
times (distributed as independent Poisson processes of unit density at each
site) to jump to an adjacent site, choosing the site on their right with some
fixed probability $p$ and that on their left with probability $q=1-p$; the
attempt succeeds if the target site is not already occupied.  For any value
of $p$ the set of extremal translation invariant stationary states is the set
of Bernoulli measures $\nu_\rho$ ($0\le\rho\le1$)~\[Ligg]; in the state
$\nu_\rho$, each lattice site is occupied independently with probability
$\rho$ and there is therefore a current $(2p-1)\rho(1-\rho)$.  The dynamics
satisfies detailed balance with respect to $\nu_\rho$ if and only if the
transitions are symmetric ($p=1/2$). 

The macroscopic mass density $u(y,t)$, which is an appropriately scaled
continuum limit of the particle density in the microscopic model, is
described~\[R]--\[Raz] by the inviscid Burgers' equation
 \begin{equation}
u_{t}+(2p-1)[u(1-u)]_y=0.\label{IB}
 \end{equation}
 It is well known~\[LPS] that solutions of \(IB) can exhibit shocks. In
the simplest example, $u(y,t)=\rho_+$ for $y>y_0(t)$ and $u(y,t)=\rho_-$
for $y<y_0(t)$, where $0\le \rho_-<\rho_+\le1$; the shock position $y_0(t)$
moves with the constant velocity $(2p-1)(1-\rho_+-\rho_-)$, as is easily
determined from conservation of mass. It is then natural to ask what this
discontinuity of the field $u$ means at the microscopic scale.

  At the microscopic level, a configuration of the ASEP at time $t$ is
fully specified by the occupation numbers $\tau_i(t)$, taking values
$0$ and $1$, of all the sites on the lattice.  To describe the profile
of the shock, we must first locate the position of the shock in each
configuration.  This is a nontrivial problem, since we have to
distinguish between the variations of density due to intrinsic
fluctuations and those due to the presence of the shock.  Fortunately
there is, for this system, a simple way of defining the position of
the shock at the microscopic level: the introduction of a single
second class particle into the system.  (Alternative and more general
methods will be discusssed in ~\[inprep].) The second class particle
acts like a hole with respect to the original, first class particles
and like a particle with respect to holes, and thus its presence does
not affect at all the dynamics of the first class particles.
Specifically, during any infinitesimal time interval $dt$, exchanges
occur on each bond as follows: 
 \begin{eqnarray}
\left. \begin{array}{c} 1 \ 0 \ \ \ \to \ \ \ 0 \ 1
\\ 2 \ 0 \ \ \ \to \ \ \ 0 \ 2 \\ 1 \ 2 \ \ \ \to \ \ \ 2 \ 1 \\
\end{array} \right\} \makebox{ with probability} \ p dt\,, \nonumber
\\
\label{hop} \\
 \left. \begin{array}{c} 0 \ 1 \ \ \  \to \ \ \  1 \ 0 \\
 0 \ 2 \ \ \  \to \ \ \  2 \ 0  \\
 2 \ 1 \ \ \  \to \ \ \  1 \ 2  \\
  \end{array} \right\} \makebox{ with probability} \ q dt\,,  \nonumber 
 \end{eqnarray}
 where a $0$, $1$ or $2$ on a site means that
this site is occupied by a hole, a regular (first class) particle or the
second class particle.

It can be shown \cite{F2} (see also \[Raz2,FK], and \[djls] for a
heuristic argument) that the velocity of the second class particle in
a uniform environment of first class particles at density $\rho$ is
$(p-q)(1-2\rho)$, so that far to the left of the shock the second
class particle moves at a velocity $(p-q)(1- 2 \rho_-)$, faster than
the shock velocity $(p-q)(1-\rho_- - \rho_+)$, whereas far to the
right of the shock the second class particle has a velocity $(p-q)(1-2
\-\rho_+)$, slower than that of the shock.  Consequently, the second
class particle is attracted to the shock and can serve as a marker for
its position.  It has in fact been proved~\cite{F2} that the second
class particle moves with velocity $(p-q)(1-\rho_--\rho_+)$
and~\cite{FKS,F2} (see also \[ABL], \[W]--\[BCFN]) that there is an
invariant measure for the system viewed from this second class
particle, in which the asymptotic densities are $\rho_\pm$.

In the present paper we describe exactly this invariant measure.  Our
approach is an extension of a matrix method which has been used in a
number of situations \cite{DEHP}--\cite{HP}, in which the weight of
each configuration is written as the matrix element of a matrix
product.  As in \cite{djls}, there are three possibilities for each
matrix in the product, $D,A$ and $E$, depending on whether the
corresponding site is occupied by a first class particle, occupied by
a second class particle, or is empty.  We show in Section~2 and
Appendix~A that when the matrices $D,A$ and $E$ satisfy certain
algebraic rules, these matrix products furnish weights for the
invariant measure we seek.  In contrast to most previous cases in
which this method was used, however, here we write probabilities of
events in the invariant measure, using matrix products, {\it directly
in the infinite system}.  This approach was used in \[DE] to recover
the results of
\cite{djls}. 

The invariant measures are parametrized by the pair of densities
$\rho_-$ and $\rho_+$.  There is a special value of the asymmetry
parameter $p$, given by $p/(1-p)=\rho_+(1-\rho_-)/\rho_-(1-\rho_+)$,
at which the measures are Bernoulli, with density $\rho_-$ to the left
of the second class particle and $\rho_+$ to its right.  This is
described in Section~3, where we also show that the algebraic rules
given in Section~2 imply certain symmetry properties of the invariant
measure.

The full expression for the shock profile, as seen from the second class
particle, is described in Section~4, and is derived in Section~5 by
constructing an explicit representation of the algebra of Section~2. 

We study in Section~6 the asymptotic behavior of this profile at large
distances from the second class particle, and show that it decays at an
exponential rate to $\rho_-$ or $\rho_+$.  The decay length is a function of
the parameters $p$ and $\rho_\pm$ which, however, becomes independent of the
asymmetry $p$ for $p/(1-p)>\sqrt{\rho_+(1-\rho_-)/\rho_-(1-\rho_+)}$. 

{}Finally, in Section~7 we show that in the weak asymmetry regime, where
$p= (1+ \epsilon )/2$ with $0<\epsilon\ll1$, the shock profile as seen
from the second class particle can be understood as the convolution of
the hyperbolic tangent profile predicted by the viscous Burgers
equation, % 
\begin{equation} u_t + [u (1-u)]_y = (1/2)u_{yy},
\end{equation} 
which is known \[DPS,KOV] to describe this regime (on
a longer time scale than that on which \(IB) holds), with the density
of the position of the second class particle in this $\tanh$ profile,
given by the derivative of the profile.

Throughout the paper, the shock will be characterized by the two asymptotic
densities $\rho_-$ to the left of the second class particle and $\rho_+$ to
the right of the second class particle, which satisfy
$0\le\rho_-<\rho_+\le1$,  and by the hopping rates $p$ to the
right and $q=1-p$ to the left, as in (\ref{hop}).   We will
often express our results in terms of $\rho_+, \rho_-, p$ and $q$ through
the parameters
 \begin{equation} 
  x={q \over p}\ , \ \ \ a= \rho_+ (1 - \rho_-) \ ,
\ \ \ b= \rho_- ( 1 - \rho_+) \ .  \label{definition}
 \end{equation}

\section{The algebra for the partially asymmetric shock measure}

In analogy with \[djls,DEHP] we write the probability of a
configuration specified by the occupation numbers $\tau_i$ ($\tau_i=0,1$)
of $m$ consecutive sites to the left of the second class particle
(which is located at the origin) and $n$ consecutive sites to its right as a
matrix element of the form
 \begin{equation}
 \langle w | \  \left\{  \prod_{i=-m}^{-1} 
   [ \tau_i D+ (1 -\tau_i) E] \right\} \   
  A \ \left\{ \prod_{j=1}^n [ \tau_j D + (1 -\tau_j) E ] \right\} 
  \   | v \rangle.
\label{weight}
\end{equation}
 Here, as in \cite{djls}, a first class particle is represented by a matrix
$D$, a hole by a matrix $E$, and the second class particle by a matrix $A$.
For example, the probability of finding the  configuration $1\;0\;1$ to the
left and $0\;1\;1\;0\;0$ to the right of the second class particle, that
is, of the configuration 
 \begin{equation}
 1 \ 0 \ 1 \ 2 \ 0 \ 1 \ 1 \ 0 \  0 \ ,\nonumber
 \end{equation}
 is given by 
 \begin{equation}
\langle w | \ D \ E \ D \ A \ E \ D^2 \ E^2 \ | v \rangle. \nonumber
 \end{equation}

In Appendix A we show, by an extension of the proof which was given in
\cite{djls}, that if the matrices $D $, $E $ and $A$ and the vectors $\langle
w|$ and $| v \rangle$ satisfy certain algebraic conditions then the weights
(\ref{weight}) are nonnegative and define an invariant measure for the ASEP,
as seen from a single second class particle, with dynamics specified by
(\ref{hop}).  The algebra is
 \begin{eqnarray}
 p D E -q E D       
  &=& (p-q) [(1- \rho_-)(1-\rho_+) D + \rho_- \rho_+ E ],
  \label{alg1} \\
 p A E  -q E A &=& (p-q)(1-\rho_-)(1-\rho_+) A ,
  \label{alg2} \\
 p D A-q A D &=& (p-q) \rho_+ \rho_- A ,
  \label{alg3}  \\
 (D + E ) | v \rangle &=& |v \rangle ,
  \label{alg4}  \\
 \langle w | (D + E ) &=& \langle w | ,
  \label{alg5} \\
 \langle w | A | v \rangle &=& 1 .
  \label{alg6}  
\end{eqnarray}
 At this stage the numbers $\rho_+$ and $\rho_-$ which appear in
(\ref{alg1})--(\ref{alg6}) are arbitrary parameters satisfying
$0\le\rho_-<\rho_+\le1$.  However, we will show in 
Section~6 that they are in fact the two asymptotic densities which
are reached as one moves away from the second class particle.

As a consequence of (\ref{weight}) and of (\ref{alg1})--(\ref{alg6}) the
microscopic profile, defined here as 
the average occupation $\langle \tau_n \rangle$ at
position $n$, is given to the right of the second class particle by 
\begin{equation}
 \langle \tau_n \rangle   = 
  \langle w | \ A \  (D+E)^{n-1} \  D \  | v \rangle , \quad\hbox{$n\ge1$},
\label{taun}
\end{equation}
 and to the left of the second class particle by
 \begin{equation}
  \langle \tau_{-n} \rangle =
  \langle w |  D \  (D+E)^{n-1}   A   | v \rangle , \quad\hbox{$n\le-1$}.
 \label{tauminusn}
 \end{equation}
 In contrast to the approach to the shock problem taken in \cite{djls},
expressions like (\ref{taun}) and \(tauminusn) are valid directly for the
infinite system; thus here we avoid completely the difficulty of taking the
infinite volume limit. 

\medskip

By an argument similar to that of Sandow~\cite{Sandow} we may verify that
(\ref{alg1})--(\ref{alg6}) suffice to determine any matrix element containing
precisely one factor of $A$. To do so we will show that any matrix
element of a product of $n+1$ operators can be reduced to a sum of matrix
elements of products of $n$ operators.  Consider, for example, a matrix
element of the form $\langle w | D O_n |v \rangle$, where $O_n$ is a
product of $n$ operators which contains a single $A$.   Then
 \begin{eqnarray}
  \langle w | D O_n |v \rangle 
  &=& x^{k+1}\langle w | O_n D |v \rangle + \hbox{l.o.t.}\nonumber\\
  &=& -x^{k+1}\langle w |  O_n E |v \rangle + \hbox{l.o.t.}\nonumber\\
  &=& -x^{n+1} \langle w | E O_n |v \rangle + \hbox{l.o.t.}\nonumber\\
  &=& x^{n+1}\langle w | D O_n |v \rangle + \hbox{l.o.t.}\ .\label{reduce}
 \end{eqnarray}
 Here l.o.t. (lower order terms) denotes matrix elements of products of $n$
matrices, $x=q / p$ as in (\ref{definition}), and $k$ is the number of
factors of $E$ in $O_n$.  To obtain \(reduce) we have used first \(alg1)
and \(alg3), then \(alg4), then \(alg1) and \(alg2), and finally \(alg5).
Since $x<1$, equation \(reduce) can be solved for
$\langle w | D O_n |v \rangle$.  This reduction permits the calculation of
matrix elements of arbitrary length.

It is important to realize, however, that the algebraic rules do not
allow us to calculate matrix elements of products of operators which
contain either no operator $A$ or more than one such operator.  For
example, $\langle w|DE|v \rangle$ or even $\langle w|v \rangle$ cannot be
calculated from these rules alone. Of course, if one has a representation
of the matrices $D$, $A$, and $E$ and of the vectors $\langle w|$ and
$|v \rangle$ which satisfies (\ref{alg1}-\ref{alg6}) then one can in
principle calculate these other matrix elements.  The values thus obtained
depend on the representation and do not appear relevant for the problem we
consider here. 

When $n$ is small the reduction described above is easy to carry out and
one can calculate directly the average occupation numbers of the sites
closest to the second class particle, as well as other simple correlation
functions.  Thus for example
\begin{eqnarray}
 \langle \tau_1 \rangle 
  &=& {  \rho_+ + \rho_- \ -  \ (1-x ) \rho_+ \rho_- \over 1+x},
\nonumber \\
 \langle \tau_{-1} \rangle 
  &=&  { x ( \rho_+ + \rho_- )\  +  \ (1-x ) \rho_+ \rho_- \over 1+x},
 \label{tau} \\
 \langle \tau_1 \tau_2 \rangle
  &=& { \rho_+^2 +  \rho_+ \rho_- + \rho_-^2 \over 1 + x+ x^2}  
    + { (-1-x+2x^2) \rho_+ \rho_- (\rho_+ + \rho_-) 
    + x (1-x)^2  (\rho_+ \rho_-)^2 \over (1+x+x^2) \ (1+x)},
 \nonumber \\
 \langle \tau_{-1} \tau_1 \rangle
   &=&  { x (\rho_+^2 +  \rho_+ \rho_- + \rho_-^2) 
   + (1-x)^2 \rho_+ \rho_- (\rho_+ + \rho_- -  \rho_+ \rho_-) \over 1+x+x^2},
 \nonumber \\
 \langle \tau_{-2} \tau_{-1} \rangle
   &=& { x^2 (\rho_+^2 +  \rho_+ \rho_- + \rho_-^2) \over 1+x+x^2}  
   +{ (2x-x^2-x^3) \rho_+ \rho_- (\rho_+ + \rho_-) 
   +  (1-x)^2  (\rho_+ \rho_-)^2 \over (1+x+x^2) \ (1+x)}.
 \nonumber 
\end{eqnarray}
In principle one could compute arbitrary correlation functions in this way,
but the calculation quickly becomes impractical as the number of sites
involved increases.  We note that, in contrast with the totally asymmetric
case $x=0$, discussed in \[djls], there are in general correlations between
the left and the right of the second class particle.  The expressions in
\(tau) are symmetric polynomial functions of $\rho_+$ and $\rho_-$.  The
symmetry follows from (\ref{alg1})--(\ref{alg3}), but we could not derive it
by an elementary argument from the dynamics. 

{}Finally, we observe that if one has a representation of $D$, $E$, $\v$, and
$\w$ which  satisfies  (\ref{alg1}), (\ref{alg4}), and  (\ref{alg5}) and for
which $\w DE-ED\v$ is finite and nonzero, then one may satisfy
(\ref{alg2})--(\ref{alg3}) by defining
\begin{equation}
A      = c \  (D E - E D),
\label{alpha}
\end{equation}
where $c$ is a constant fixed by the condition (\ref{alg6}).  The
representation constructed in Section~5 is obtained in this way.  We must
emphasize, however, that there are representations of interest in which
\(alpha) is not satisfied.  One such representation is used in Section~3
below to study the system at a special value $x^*$ of $x$; note also that
if one wished to use the algebra (\ref{alg1})--(\ref{alg6}) when $x=1$ then
\(alg1) would imply that $D$ and $E$ commute, so that \(alpha) would be
inconsistent with the normalization \(alg6).  More generally,
(\ref{alg4}) and (\ref{alg5}) imply that
$\langle w | DE | v \rangle = \langle w | ED | v \rangle $, so that
$\langle w | DE-ED | v \rangle $ can be nonzero only if
$\langle w | DE | v \rangle $ and $\langle w | ED | v \rangle $ are
infinite. Thus finite dimensional representations of the algebra cannot
satisfy \(alpha); the representation constructed in Section~5 is, as
expected from this remark, infinite dimensional.

\section{Elementary consequences of the algebra}

We present here two simple consequences of the representation of
the invariant measure described in Section~2. 

{}First we note that there is a special value $x^*$ of the ratio $x=q/p$ for
which the measure becomes a Bernoulli measure with density $\rho_+$ to the
right of the second class particle and density $\rho_-$ to the left
of the second class particle:
\begin{equation}
x^* = {\rho_- (1- \rho_+) \over \rho_+ ( 1- \rho_-) } = {b\over a}\ .
\label{xstar}
\end{equation}     
This can be seen by verifying that when $x=x^*$ the following formulas define
a two-dimensional representation the algebra (\ref{alg1})--(\ref{alg6}):
\begin{eqnarray}
D  = \left( \begin{array}{cc} \rho_+&0 \\ 0& \rho_-\\ \end{array} \right)
   \  ,  \ \ 
E = \left( \begin{array}{cc} 1-\rho_+&0 \\ 0& 1-\rho_-\\ \end{array} \right)
   \  ,  \ \ 
A = \left( \begin{array}{cc} 0&0 \\ 1& 0 \\ \end{array} \right)
  \  ; && \nonumber\\ 
 \langle w | = (0,1) \ ,  \ \ 
  | v \rangle = \left( \begin{array}{c} 1 \\  0 \\ \end{array} \right)\ .
  \hskip100pt &&
  \label{xstar1}
\end{eqnarray}
 It is also easy to check directly that this Bernoulli measure is
stationary.  When $x=x^*$ the profile on both sides of the second class
particle is flat ($\langle \tau_n \rangle = \rho_+$ and
$\langle \tau_{-n} \rangle = \rho_-$ for all $n \geq 1$) and the occupation
numbers at all sites are independent:
 \begin{equation}
\E{\tau_{-m}\ldots\tau_{-1}\tau_1\ldots\tau_n} = 
   \E{\tau_{-m}}\ldots\E{\tau_{-1}}\E{\tau_1}\ldots\E{\tau_n}
 = \rho_-^m\rho_+^n.
    \end{equation}
 The value (\ref{xstar}) of $x^*$ plays the role of a disorder line in
equilibrium statistical mechanics models \cite{STE,RUJ}.

Second, we obtain a number of identities satisfied by the invariant measure
for all $x$, $\rho_-$, and $\rho_+$.  
These will be derived under the assumption that $A$ has the special form
\(alpha), but this is not a restriction on their validity, since we know that
there exists at least one representation, constructed in Section~5, for which
this is true.  The simplest of these identities is
 \begin{eqnarray}
 \langle \tau_n \rangle -\langle \tau_{n+1} \rangle 
  &=& \langle w |  A (D + E )^{n-1} (D E - E D )| v \rangle\nonumber\\
  &=& \langle w | (D E - E D ) (D + E )^{n-1} A | v \rangle\label{id1}\\
 &=& \langle \tau_{-n-1} \rangle -\langle \tau_{-n} \rangle\ ,\nonumber
 \end{eqnarray}
 which implies that the shock profile has the symmetry
 \begin{equation}
 \langle \tau_n \rangle + \langle \tau_{-n} \rangle =
 \langle \tau_{n+1} \rangle + \langle \tau_{-n-1} \rangle  \ .
 \label{sym0}
 \end{equation}
 {}From  the known results (\ref{tau}) for the first sites  it follows that
the common value  is
\begin{equation}
\langle \tau_n \rangle + \langle \tau_{-n} \rangle = \rho_+ + \rho_- \ .
\label{sym}
\end{equation}
Note that the right hand side of (\ref{sym}) is consistent with the 
asymptotic values $\langle \tau_n \rangle \to \rho_+ $ and
$\langle \tau_{-n} \rangle \to \rho_- $ as $n \to \infty$.
                      
One can obtain in the same manner identities involving higher correlations. For
example, from (\ref{alpha}) one has
\begin{eqnarray}
  \langle w | A (D + E )^{n'} (D E - E D) (D + E )^{m'} D | v \rangle 
  = \langle w | (D E - E D) 
  (D+ E  )^{n'}  A    (D + E  )^{m'}  D   | v \rangle,
\nonumber \\
\langle w | A (D+ E )^{n'} D (D + E )^{m'} (D E   - E   D) | v \rangle 
 = \langle w | (D E - E D) (D+ E )^{n'} D (D + E )^{m'} A | v \rangle,
\nonumber \\
\langle w | D (D+ E )^{n'} A (D + E  )^{m'} (D E - E D) | v \rangle 
 = \langle w | D (D+ E )^{n'} (D E - E D) (D + E )^{m'} A | v \rangle,
\nonumber\\
\end{eqnarray}
and by choosing $n'=n-1$ and $m'=m-1$ one obtains symmetry relations
involving pair correlation functions
\begin{eqnarray}
  \langle \tau_{n} \tau_{n+m+1} \rangle 
   - \langle \tau_{n+1} \tau_{n+m+1} \rangle 
 &=& \langle \tau_{-n-1} \tau_{m} \rangle
   - \langle \tau_{- n} \tau_{m} \rangle, \nonumber \\
   \langle \tau_{n} \tau_{n+m} \rangle 
   - \langle \tau_{n} \tau_{n+m+1} \rangle 
  &=& \langle \tau_{-n-m-1} \tau_{-m} \rangle 
   - \langle \tau_{-n-m} \tau_{-m} \rangle, \label{rel2} \\
  \langle \tau_{-n} \tau_{m} \rangle 
   - \langle \tau_{-n} \tau_{m+1} \rangle 
 &=& \langle \tau_{-n-m-1} \tau_{-m-1} \rangle 
  - \langle \tau_{-n-m-1} \tau_{-m} \rangle, \nonumber
\end{eqnarray}
which, through linear combinations, lead to the fact that 
 $\langle \tau_n \tau_{n+m}  \rangle +
\langle \tau_{-n}  \tau_{m}  \rangle +
\langle \tau_{-n-m}  \tau_{-m}  \rangle$
does not depend on $n$ or $m$ and therefore is given from (\ref{tau}) by
\begin{equation}
\langle \tau_n \tau_{n+m}  \rangle +
\langle \tau_{-n}  \tau_{m}  \rangle +
\langle \tau_{-n-m}  \tau_{-m}  \rangle = \rho_+^2 + \rho_+ \rho_- + \rho_-^2 
\label{identity}
\end{equation}
Note that this common value is again consistent with the asymptotic
densities $\rho_\pm$ and the independence of the occupation numbers
$\tau_{\pm n}$, $\tau_{\pm m}$ as $n,m\to\infty$ (see Remark~6.1). 

\section{The expression for the density profile}

Here we present an expression for the profile
$\langle\tau_n\rangle$, based on an explicit representation of the algebra
(\ref{alg1})--(\ref{alg6}) which will be constructed in Section~5.  The
formula for the profile to the right of the second class particle is
\begin{equation}
\langle \tau_{n} \rangle = \rho_+  + \sum_{k=-\infty}^\infty P_{n-1}(k) F_k\;,
   \qquad\hbox{for $n\ge1$,}
\label{exact}
\end{equation}
where the  $P_n(k)$ are defined by the recursion
\begin{eqnarray}
P_0(k) &=& \delta_{k,0},
\label{P0}\\
P_{n+1}(k) &=& a P_n(k-1) + (1-a-b) P_n(k) + b P_n(k+1) ,
\label{Pn}
\end{eqnarray}
and $F_k$ is given by
\begin{equation}
F_k = {1\over a-b}
 (a ^2 f_{k+2} -a(a+b) f_{k+1} + b(a+b) f_{k-1} - b^2 f_{k-2}) ,
\label{Fk}
\end{equation}
with
\begin{equation}
f_k = k \ {x^k \over 1 - x^k} ,\qquad\hbox{for $k\ne 0$.}
\label{fk}
 \end{equation} 
 $f_0 $ can be defined arbitrarily because its contribution to
(\ref{exact}) cancels out: indeed, it is easy to check that
(\ref{P0}) and (\ref{Pn}) imply that $a^kP_n(-k)= b^k P_n(k)$ for all $k$, so
that the coefficient of $f_0$ in (\ref{exact}) is always zero.  (We will
introduce below a particular value of $f_0$ which is convenient in
intermediate computations.)  The profile to the left of the second class
particle---$\E{\tau_n}$ for $n\le-1$---can easily be obtained from the
symmetry (\ref{sym}).

 More complicated correlation functions have similar
expressions, which can be derived in the same way  or, in some cases,
directly from \(exact) and the algebra.  For example, from
(\ref{alg1})--(\ref{alg6}), one can show that
 \begin{equation}
(1-x) D^2 |v \rangle = [ D (D+E) - x (D+E)D - (1-x)(1-\rho_+ - \rho_-) D -
(1-x) \rho_+ \rho_- ] | v \rangle,
 \end{equation}
  and this implies that 
 \begin{equation}
 (1-x) \langle \tau_n \tau_{n+1} \rangle 
  = \langle \tau_n \rangle - x \langle \tau_{n+1} \rangle -
  (1-x) (1 - \rho_+ - \rho_-) \langle \tau_n \rangle - (1-x) \rho_+ \rho_- ,
 \end{equation}
 which gives, using (\ref{exact}),
 \begin{equation}
\langle \tau_n \tau_{n+1} \rangle =
 \rho_+^2  + \sum_{k=-\infty}^\infty P_{n-1}(k)  \ G_k \ ,
 \end{equation}
 where
 \begin{equation}
G_k =(\rho_+ + \rho_-) F_k 
   + {x \over 1 - x} [(a+b)F_k- a F_{k+1} -b F_{k-1}]\ .
 \end{equation}

 \medskip\noindent
 {\bf Remark 4.1:} The quantity $P_n(k)$ given by (\ref{P0}) and (\ref{Pn})
can be interpreted as the probability of finding a biased random walker on
site $k$ at time $n$, given that it was at the origin at time 0.  We did not
find a simple physical interpretation for the presence of this random walk in
the expression (\ref{exact}) for the profile but we believe that
understanding the origin of this biased random walk would give a better
insight in the whole problem of the description of a shock as seen from a
second class particle. 

\section{Representation of the algebra}

We now describe the explicit representation of the algebra
(\ref{alg1})--(\ref{alg6}) used to derive the formulas of Section~4 for the
profile and to study the weak asymmetric limit in Section~7.  The
operators $D$, $E$, and $A$ will act on an infinite dimensional
vector space with basis $\{\,|n\rangle\mid n=0,\pm1,\pm2,\ldots\,\}$.  Let us
define the left shift operator $L$, the two diagonal operators $S$ and $T$,
and a vector $|v\rangle$ and dual vector $\langle w|$ by
 \begin{eqnarray}
L\,|n\rangle&=&|n-1\rangle\;,\\
T\,|n\rangle&=&x^n(1+x^n)^{-1}|n\rangle\;,\\
S\,|n\rangle&=&\Bigl[x^n(1+x^n)^{-1}-x^{n-1}(1+x^{n-1})^{-1}
    \Bigr]|n\rangle\;,\\
 |v\rangle &=& \sum_{n=-\infty}^\infty |n\rangle\;,\qquad\qquad
 \langle w| = \sum_{n=-\infty}^\infty \langle n|\;.\label{vwA}
 \end{eqnarray}
These are easily seen to satisfy the following relations:
 \begin{eqnarray}
 &pLT-qTL = (p-q)TLT, \qquad
     pTL^{-1}-qL^{-1}T = (p-q)TL^{-1}T;&\label{QCR}\\
&(p-q)T(aL-bL^{-1})T
 = paLT-qaTL-pbTL^{-1}+qbL^{-1}T;&\label{QCR2}\\
 &TL-LT = -LS, \qquad\qquad
     L^{-1}T-TL^{-1} = -SL^{-1};&\label{QCR3}\\
 &L|v\rangle =|v\rangle,\qquad\qquad  \langle w|L = \langle w|.&
 \end{eqnarray}

  Next we define the operators of our representation:
 \begin{eqnarray}
 D &=& \rho_-\rho_++aL-(\a L-\b)T(\a+\b L^{-1}),\label{delA}\\
 E &=& (1-\rho_-)(1-\rho_+)+bL^{-1}+(\a L-\b)T(\a+\b L^{-1}),\label{epsA}\\
 A &=&-(a-b)^{-2}(DE-ED).\label{alphA}
 \end{eqnarray}
 Then the formula (\ref{alg1}) for $pDE-qED$
follows immediately from (\ref{QCR2}), and as noted in Section~2 the
relations (\ref{alg2})--(\ref{alg3}) for $pAE-qEA$
and $pDA-qAD$ are automatically satisfied
by (\ref{alphA}). Moreover
 \begin{equation}
 D + E = aL+(1-a-b)I+bL^{-1},\label{dpe}
 \end{equation}
  so that $(D+E)\v=\v$ and $\w(D+E)=\w$.  Finally,
(\ref{QCR3}) implies that
 \begin{equation}
 A = -(a-b)^{-2}(\a L-\b)(aLS-bSL^{-1})(\a+\b L^{-1}),\label{alphB}
 \end{equation}
and from $\langle w|S|v\rangle=-1$ it follows that 
$\langle w|A |v\rangle=1$.

Note that, as expected from the remark following (\ref{alpha}), the matrix
elements $\langle w |v \rangle $, $\langle w | DE | v \rangle $ and
 $\langle w | ED |v \rangle$ are all infinite.  In this representation the
presence of a factor $A$ in a matrix product---or more specifically, that of
$S$ (see \(alphB)), a diagonal matrix whose diagonal elements $S_{jj}$
decrease exponentially fast to $0$ as $j\to\pm\infty$---renders the
 $\w\;\cdot\;\v$ matrix element of that product finite. 

 We now turn to the evaluation of $\langle\tau_n\rangle$.  First we observe
that, for $k>0$,
 \begin{eqnarray}
\sum_{j=-N}^N{x^{j-k}\over1+x^{j-k}}{x^j\over1+x^j}
  &=&\sum_{j=-N}^N{1\over1-x^k}
    \left({x^j\over1+x^j}-x^k{x^{j-k}\over1+x^{j-k}}\right)\nonumber\\
  &=&\sum_{j=-N}^N{x^j\over1+x^j}
   +{x^k\over1-x^k}\left(\sum_{j=N-k+1}^N{x^j\over1+x^j}
    -\sum_{j=-N-k}^{-N-1}{x^j\over1+x^j}\right).\nonumber\\
  \label{sum}
 \end{eqnarray}
 {}From (\ref{sum}) we have, for $k>0$,
 \begin{eqnarray}
 \langle w|SL^kT|v\rangle 
  &=&\lim_{N\to\infty} \sum_{j=-N}^N
    \left({x^{j-k}\over1+x^{j-k}}-{x^{j-1-k}\over1+x^{j-1-k}}\right)
     {x^j\over1+x^j}\nonumber\\
   &=&\lim_{N\to\infty}\left[
  {x^k\over1-x^k}\left(\sum_{j=N-k+1}^N{x^j\over1+x^j}
    -\sum_{j=-N-k}^{-N-1}{x^j\over1+x^j}\right)\right.\nonumber\\
 &&\qquad\qquad -\left.{x^{k+1}\over1-x^{k+1}}
     \left(\sum_{j=N-k}^N{x^j\over1+x^j}
    -\sum_{j=-N-k-1}^{-N-1}{x^j\over1+x^j}\right)\right]\nonumber\\
  &=& f_{k+1} - f_k,
\label{ftmp}
 \end{eqnarray}
 with $f_k=kx^k/(1-x^k)$ as in (\ref{fk}).  From the easily verified identity
 \begin{equation}
   \langle w|SL^jT|v\rangle +\langle w|SL^{-j-1}T|v\rangle = -1,\label{iden}
 \end{equation}
 it then follows that (\ref{ftmp}) also holds for $k<-1$, with $f_k$ again
given by (\ref{fk}).  Now let us define $f_0=f_1- \langle w|ST|v\rangle$; then
(\ref{ftmp}) holds for $k=0$ and also $k=-1$ (using (\ref{iden}) again), and
thus for all $k$.  It is in fact not necessary to evaluate $f_0$, by the
remark following \(fk).

Using  \(delA), (\ref{alphB}), and (\ref{ftmp}) we have, for all $k$,
 \begin{eqnarray}
  \langle w|A      L^k D    |v\rangle
  &=&\rho_-\rho_++a
   + (a-b)^{-1}\langle w|(aLS-bSL^{-1}) L^k(aL-bL^{-1})T|v\rangle\nonumber\\
  &=& \rho_+ + (a-b)^{-1}\langle w|S(a^2L-ab-abL^{-1}+b^2L^{-2})L^kT
         |v\rangle\nonumber\\
  &=& \rho_+ + (a-b)^{-1}\Bigl(a^2f_{k+2}-a(a+b)f_{k+1}
   +b(a+b)f_{k-1}-b^2f_{k-2}\Bigr).\hskip15pt
 \end{eqnarray}
 The formula (\ref{dpe}) for $ D + E $ yields by induction
 \begin{equation}
 ( D + E  )^n=\sum_kP_n(k)L^k 
 \end{equation}
 for $n\ge 0$, where the $P_n(k)$ are defined by (\ref{P0}) and (\ref{Pn}). 
Thus for $n\ge1$,
 \begin{eqnarray}
 \langle \tau_{n} \rangle 
    &=& \langle w|A     ( D    + E      )^{n-1} D    |v\rangle\nonumber\\
  &=&  \sum_{k=-\infty}^\infty P_{n-1}(k) 
 \langle w|A      L^k D    |v\rangle   \label{tn1} \\
 &=& \rho_+  + {1\over a-b}\sum_{k=-\infty}^\infty P_{n-1}(k)
   \Bigl(a^2f_{k+2}-a(a+b)f_{k+1}
   +b(a+b)f_{k-1}-b^2f_{k-2}\Bigr),\nonumber
 \end{eqnarray}
and this completes the derivation of the formula (\ref{exact}) for
$\langle\tau_{n}\rangle$. 
  
 {}From the expression (\ref{exact}) it is possible to evaluate various
asymptotics of the profile.  The two main limits that one might wish to
consider are the limit $n \to \infty$  describing the tail of the profile,
discussed in the following section (where we also comment briefly on the
$\rho_+\searrow\rho_-$ limit), and the limit of a weak asymmetry, $ x
\to 1$, discussed in Section~7. 

\section{Spatial asymptotics of the density profile}

 We now discuss the behavior of the profile $\E{\tau_n}$ at large $n$,
basing our discussion on \(exact).  Because $a>b$, the biased random walker
governed by (\ref{Pn}) goes to $+\infty$ as $n \to \infty$, with
distribution $P_n(k)$ concentrated around $k\simeq n(a-b)$.  Moreover,
$F_k$ approaches $0$ as $k$ approaches $\infty$, decaying exponentially (as
$kx^k$), so that the sum in (\ref{exact}) vanishes in the $n\to\infty$
limit and thus the asymptotic density on the right is $\rho_+$, i.e., 
 \begin{equation}
\langle \tau_n \rangle \to \rho_+ \qquad \hbox{as $n\to\infty$.}
 \end{equation}
 {}From the
symmetry (\ref{sym}), the asymptotic density at the left is $\rho_-$. 

To obtain the approach of $\E{\tau_n}$ to its $n \to \infty$ limit,
we need to estimate how the sum on the right hand side of (\ref{exact})
vanishes.  As observed above, $F_k$ decays exponentially as $k\to\infty$, and
it is easy to check that it approaches the constant value $-(a-b)$ as
$k\to-\infty$.  Because of this behavior there are two cases to consider: the
sum is dominated either by large values of $k$ or by values of $k$ close to
0, depending on the value of $x$. 

\medskip
\noindent{\bf Case I: $x >  (b/a)^{1/2}$} 

 \smallskip
 We first explore the consequences of assuming that the sum in
(\ref{exact}) is dominated by large values of $k$.  If this is so, one can
use the approximation
 \begin{equation}
f_k \simeq k   x^k  ,
 \end{equation}
 and then the identity
\begin{equation}
\sum_{k=-\infty}^\infty x^k P_n(k) = \left( 1-a-b+ax+{b\over x} \right)^n .
\label{SUM}
\end{equation}
 This leads to
\begin{eqnarray}
\langle \tau_{n} \rangle - \rho_+ \simeq 
(n-1) { \left( ax - {b/x} \right)^2 (ax-a-b+{b/x} ) \over a-b} 
   \left( 1-a-b+ax+{b\over x} \right)^{n-2}  \nonumber \\
 +{ 2 a^2 x^2- (a+b) a x - (a+b) {b/x} + 2 b^2/x^2 \over a-b}
 \left(1-a-b+ax+{b\over x} \right)^{n-1} .\label{case1}
\end{eqnarray}
 The values of $k$ which dominate the sum (\ref{SUM}) are
 $k\simeq n (ax-b/x)/(1-a-b+ax+b/x)$, so that the assumption that large
values of $k$ dominate the sum in \(exact) would be inconsistent if $x^2 <
b/a$.  Conversely, when $x^2 > b/a$, \(case1) gives the asymptotic behavior
of $\E{\tau_n}$. 

\medskip
\noindent{\bf Case II: $x <  (b/a)^{1/2}$} 

\smallskip
Since $F_k$ is constant for $k\to-\infty$ and $P_n(k)$ increases with $k$ for
$k<0$, it is clear that the sum over $k$ must be dominated by values of $k$
near $k=0$ (i.e. which do not scale with $n$).  It is convenient to use the
following expression for the solution $P_n(k)$ of (\ref{Pn}):
 \begin{equation}
  P_n(k) = \sum_{m=0}^n {n \choose m} (1-a-b)^{n-m} 
  {m \choose {m+k \over 2} } b^{m-k \over 2} a^{m+k \over 2} .
 \end{equation}
  {}For large $m$ and for $k$ of order $1$, one can write
 \begin{eqnarray}
{m \choose {m+k \over 2}} 
   &\simeq& 2^m \  \sqrt{2 \over \pi  m} \ \left( 1 - {k^2 \over 2 m}\right),
\ \ \ \makebox{for $m+k$ even,} \nonumber\\
 {m \choose {m+k \over 2}} &=& 0,  \ \ \ \makebox{for $m+k$ odd.} 
 \end{eqnarray}
  Then
 \begin{eqnarray}
  \langle \tau_{n} \rangle - \rho_+ 
  &\simeq& {1\over2}\sqrt{2\over \pi} \sum_{m=0}^{n-1} {n-1 \choose m} 
  (1-a-b)^{n-m-1} \left( ab \right)^{m/2} {2^m\over m^{1/2}}
  \sum_{k=-\infty}^\infty \left(a\over b\right)^{k/2}
   \left(1-{k^2\over2m}\right)F_k\nonumber\\
  &\simeq& {1 \over 2} \sqrt{2\over \pi} \sum_{m=0}^{n-1} {n-1 \choose m} 
  (1-a-b)^{n-m-1}   2^m \left( ab \right)^{m/2} 
  {Z \over 2 m^{3/2}}.\label{diff1}
 \end{eqnarray}
 Here we have used the fact that $\sum_{k=-\infty}^\infty (a/b)^{k/2}F_k=0$
(see \(Fk) and \(fk)) and set 
 \begin{equation}
Z=-\sum_{k=-\infty}^\infty k^2\left(a\over b\right)^{k/2}F_k
  = -{4 \sqrt{ab} \ (\sqrt{a} - \sqrt{b})^2 \over a-b} 
    \sum_{k=-\infty}^\infty 
    { k^2 x^k \over 1 - x^k} \ \left( a \over b \right)^{k/2} .\label{Z}
 \end{equation}
 The extra factor of $1/2$ in \(diff1) comes from the fact that we have
replaced a sum in which $m$ and $k$ have the same parity by a sum over all
$m$ and $k$.  Finally, using the fact that
 $\sum_m {n \choose m} x^{n-m} y^m m^{-\alpha} \simeq (x+y)^{n+\alpha} 
 (ny)^{-\alpha}$, since this sum is dominated by values of $m$ near
$ny/(x+y)$, \(diff1) becomes
 \begin{equation}
  \langle \tau_{n} \rangle - \rho_+ 
  \simeq  {Z \over  8  \sqrt{  \pi } \ (ab)^{3/4} }   
  \ {1 \over n^{3/2}} \  [ 1 - (\sqrt{a}- \sqrt{b})^2]^{n+{1 \over 2}} .
 \nonumber
 \end{equation}

 By pairing the $\pm k$ terms in \(Z) we see that if $x<x^*=b/a$ then $Z>0$
and asymptotically $\E{\tau_n}>\rho_+$, while if $x>x^*$, asymptotically
$\E{\tau_n}<\rho_+$.  Numerical computation indicates that $\E{\tau_n}-\rho_+$
has the same sign for all $n$ and is in fact monotonically decreasing for
$x<x^*$ and monotonically 
increasing for $x>x^*$.  Note also that $Z$ vanishes for $x=x^*$, as
expected from the special nature of the measure at this value of $x$ (see
Section~3).

   When $x=0$, \(Z) becomes
 \begin{equation}
 Z= {4 ab \over ( \sqrt{a} - \sqrt{b})^2} ,\nonumber
 \end{equation}
 leading  to
 \begin{equation}
  \langle \tau_{n} \rangle - \rho_+ 
  \simeq  {(ab)^{1/4} \over  2   \sqrt{\pi}  (\sqrt{a} - \sqrt{b})^2   }   
   \ {1 \over n^{3/2}} \  [ 1 - (\sqrt{a}- \sqrt{b})^2]^{n+{1 \over 2}}.
  \nonumber
 \end{equation}
 This agrees  with (6.26) of \[djls] (and with (7.6) of that paper after
the correction of a misprint: the exponent $2i-1$ should be $2i+1$).

By a modification of the above calculation one may also find the asymptotics
when $\rho_+=\rho_-$.  The result is independent of $x$ and is that given in
\[djls]: $\E{\tau_n}\simeq\rho+\sqrt{\rho(1-\rho)/\pi n}$ when $n\gg1$, where
$\rho=\rho_+=\rho_-$. 

The various asymptotics for the profile derived here are summarized
in Figure 1.  Figure 2 shows the profile for some typical parameter values. 

 \medskip\noindent
  {\bf Remark 6.1} Using the explicit representation of the algebra
developed in Section~5 it  can be verified that the measure is
asymptotically Bernoulli as $n\to\pm\infty$, i.e., that if $\Gamma$ is any
operator product containing  $j$ factors of $D$ and $k$ factors of $E$ then 
 \begin{equation}
\lim_{n\to\infty}\Ewv{A(D+E)^n\Gamma}=\rho_+^j(1-\rho_+)^k,
 \end{equation}
 with a similar limit at $-\infty$.  The approach is exponentially fast, with
the same decay rate as for the profile, that is, $(1-a-b+ax+b/x)^n$ when
$x>(b/a)^{1/2}$ and $[1-(\sqrt a-\sqrt b)^2)]^n$ when $x<(b/a)^{1/2}$.  More
generally, if $\Delta$ is a product of $D$'s, $E$'s and precisely one $A$,
then $\Ewv{\Delta(D+E)^n\Gamma}\to\Ewv{\Delta}\rho_+^j(1-\rho_+)^k$ as
$n\to\infty$, with the same rate of approach (and again a similar limit at
$-\infty$).

\section{The weakly asymmetric limit}

 In this section we study the limiting behavior of the model as the asymmetry
$\epsilon=p-q$ becomes small, with $\rho_\pm$ fixed.  In this case the
profile $\E{\tau_n}$ becomes very broad and as $\epsilon\to0$ depends only 
on the scaled variable $y=n\epsilon$.  To verify
this, note that for $\epsilon$ small and $k$ large it follows from 
 \begin{equation}
    x=q/p=1-2\epsilon+O(\epsilon^2)
 \end{equation}
  that 
 \begin{equation}
  f_{k+p} = (k+p){x^{k+p}\over1-x^{k+p}}
  \simeq  k { x^k \over 1 - x^k} + p  { x^k \over 1 - x^k} 
       - 2 p k \epsilon {x^k \over (1- x^k)^2},
 \end{equation}
 so that 
 \begin{equation}
F_k \simeq (a-b)\left[{ x^k \over 1 - x^k} 
       - 2 k \epsilon {x^k \over (1- x^k)^2}\right]\label{Fkaprx}.
 \end{equation}
 Now for large $n$ the probability $P_n(k)$ is concentrated around $k = n
(a-b)$; if we introduce the variable $y= n\epsilon$, use the
approximation \(Fkaprx), and set $k=(a-b)y/\epsilon$ we obtain
 \begin{eqnarray}
  \E{\tau_n}
   &\simeq& \rho_+  
    +\lambda\left [ {2 e^{-4 \lambda y} \over 1 - e^{- 4 \lambda y} } 
    - \lambda y {8 e^{- 4 \lambda y} \over (1 - e^{- 4 \lambda y})^2} \right] 
   \nonumber\\
   &\simeq& {\rho_+ + \rho_- \over 2} +
    \lambda \left[ \coth 2\lambda y 
     - {2\lambda y \over \sinh^2 2\lambda y } \right],
   \label{scaling}
 \end{eqnarray}
where $\lambda = (a-b)/2=(\rho_+-\rho_-)/2$.  It is easy to see that the
terms neglected in \(scaling) are $O(\epsilon)$ when $y$ is of order 1.
Note that the asymptotic behavior of (\ref{scaling}) agrees with the
$x \to 1$ limit of \(case1).

 This scaling form for the profile can be understood in terms of the
viscous Burgers equation for a hydrodynamic density variable $u(y,s)$:
 \begin{equation}
  u_s
  + [u(1-u)]_y
  = (1/2)u_{yy}.
  \label{Bur}
 \end{equation}
 It is shown in \[DPS], \[KOV] (see also \[LOB]) that the weakly asymmetric
macroscopic limit of the ASEP, in scaled variables $y=n\epsilon$ as above and
$s$, related to the microscopic time $t$ by $s=\epsilon^2 t$, is indeed
\(Bur).  For example, if the initial state of the particle system with
parameter $\epsilon$ is a product measure with density $u_0(\epsilon n)$ at
site $n$, then the observed density $u(y,s)$ at position $\epsilon^{-1}y$ and
time $\epsilon^{-2}s$ is the solution of (\ref{Bur}) with initial condition
$u(y,0)=u_0(y)$.  Equation (\ref{Bur}) is diffusive and does not exhibit
discontinuous shocks; instead of the jump in density from $\rho_-$ to
$\rho_+$ we now have a smooth profile $u(y,s)=\rho(y-vs)$, traveling with
velocity $v=1-\rho_+-\rho_-$, with
 \begin{equation} \rho(y)=(\rho_-+\rho_+)/2
  +  \lambda\tanh2\lambda y,
 \label{tanh}
 \end{equation}
 and $\lambda=(\rho_+-\rho_-)/2$.  Thus the shock width on the microscopic
scale diverges in this limit as $\epsilon^{-1}$. 

 We can now interpret \(scaling), after putting $y=\epsilon n$, as the
shock profile seen from a second class particle which does not have a fixed
location relative to the tanh profile; instead, its position is distributed
in such a way that it sees all densities between $\rho_-$ and $\rho_+$ with
equal probability. Then  the probability of finding the second class particle
between $y$ and $y+dy$ is $d\rho(y)=Q(y)\,dy$, where
 \begin{equation}
 Q(y) = {1\over\rho_+-\rho_-}\rho'(y) = \lambda\cosh^{-2}2\lambda y,
 \label{dens}
 \end{equation}
 and the limit of the one particle density function will be
 \begin{equation}
 \lim_{\epsilon\to0}\langle\tau_{\lfloor\epsilon^{-1}y\rfloor}\rangle
 = \int_{-\infty}^{\infty}\rho(y+z)Q(z)\,dz
  =  {\rho_+ + \rho_- \over 2} + 
  \lambda\left[\coth\lambda y - {2\lambda y\over \sinh^2 2\lambda y } \right],
 \label{one}
 \end{equation}
 in agreement with \(scaling); here $\lfloor z\rfloor$ denotes the greatest
integer $k$ satisfying $k\le z$.

 \medskip\noindent
 {\bf Remark 7.1:} In \[FKS] the shock is viewed, not from the position of a
second class particle, but from a location defined in a rather less direct
manner.  With that definition the profile seen in the weakly asymmetric limit
is $\rho(y)$, not the convolution \(one). 
 \medskip
 
There are various possible direct justifications for the density \(dens).  A
heuristic argument may be based on a comparison of two systems: in one of
these the second class particle is treated as a first class particle and in
the other as a hole. In the latter case the shock position, however
determined, is effectively translated a macroscopic 
distance $\epsilon/(\rho_+-\rho_-)$ relative to its position in the
former.  Thus the density of the second class particle is
$\bigl(\rho(y+\epsilon/(\rho_+-\rho_-))-\rho(y)\bigr)/\epsilon\simeq
\rho'(y)/(\rho_+-\rho_-)$.  A rigorous version of this argument is given in
\[F2]. 

The above picture is confirmed if we compute other aspects of the
$\epsilon\to0$ limiting behavior of the entire invariant measure.  Consider
first the distribution of occupation numbers at a finite number of sites
with both microscopic and macroscopic spacing.  To give a concrete example
we consider
 \begin{equation}
 \Ewv{A(D+E)^{n_1}DED
   (D+E)^{n_2-n_1}EEDD}
 \end{equation}
 where $n_1=\lfloor y_1/\epsilon\rfloor$ and $n_2=\lfloor
y_2/\epsilon\rfloor$ for some macroscopic positions $y_1$, $y_2$ with
$0<y_1<y_2$; this is the probability that of three specific sites at
$y_1$ the first and third are occupied and the second
empty, and that the first two of four sites at $y_2$ are empty and the second
two occupied.  Then
 \begin{eqnarray}
 && \hskip-30pt\lim_{\epsilon\to0}\;\Ewv{A(D+E)^{n_1}DED
   (D+E)^{n_2-n_1}EEDD}\nonumber\\
 && \hskip20pt=\;\int_{-\infty}^{\infty}
    \rho(y_1+z)^2\bigl(1-\rho(y_1+z)\bigr)
  \rho(y_2+z)^2\bigl(1-\rho(y_2+z)\bigr)^2\;Q(z)\,dz.
 \label{7sites}
 \end{eqnarray}
 We may think of the right hand side of \(7sites) as arising from a
distribution of particles in which sites at macroscopic positions $y$ are
occupied with probability $\rho(y)$ and all occupation numbers are
independent; this distribution is viewed from a random position $z$ (the
position of the second class particle) distributed according to $Q(z)\,dz$. 
Equation \(7sites)---in fact, its generalization to an arbitrary number of
sites---will be proved in Appendix~B; in particular, this furnishes an
alternate derivation of the special case \(scaling) of one site. 

We can also show that, in a certain sense, the typical configuration in the
$\epsilon\to0$ limit has on the macroscopic scale the shape of the tanh
profile \(tanh).  To do so we introduce a coarse graining on an intermediate
scale $\epsilon^{-\beta}$, where $0<\beta<1$: for each macroscopic position
$y$ we define a random variable $B_\epsilon(y)$, the empirical density in a
block of size $\epsilon^{-\beta}$ at $y$, by
 \begin{equation}
  B_\epsilon(y)=\epsilon^\beta\sum_{k=1}^{\lfloor \epsilon^{-\beta}\rfloor}
    \tau_{\lfloor y/\epsilon\rfloor+k}.\label{Bdef}
 \end{equation}
 Now consider, for very small $\epsilon$, a configuration of the system which
after coarse graining looks like the tanh profile \(tanh) on the macroscopic
scale.  The second class particle will be located at some macroscopic
position $z$ relative to the center of this tanh profile and hence will see
locally a density $\rho(z)$; thus $B_\epsilon(0)\simeq\rho(z)$ and
$B_\epsilon(0)$ determines the position $z$ by
$z\simeq\rho^{-1}(B_\epsilon(0))$.  At position $y$ relative to the second
class particle the density will be $\rho(z+y)$, and we are led to expect a
strong correlation between the empirical densities at positions $0$ and $y$
relative to the second class particle, given by
 \begin{equation}
B_\epsilon(y)\simeq \rho\bigl(y+\rho^{-1}\bigl(B_\epsilon(0)\bigr)\bigr).
 \label{corr}
 \end{equation}
 We show in Appendix~B that \(corr) holds with arbitrary accuracy, and with
probability arbitrarily close to one, for sufficiently small
$\epsilon$, that is, that for any $\eta>0$,
 \begin{equation}
  \lim_{\epsilon\to0}\;
 \Pr\left[B_\epsilon(y)-\rho\bigl(y+\rho^{-1}\bigl(B_\epsilon(0)\bigr)\bigr)
 > \eta\right] = 0\label{eta}
  \end{equation}
 (convergence in probability).  Equation \(eta) essentially says that, with
probability one, the coarse grained densities in the $\epsilon\to0$ limit lie
on some translate of the profile $\rho(x)$.

\section{Concluding remarks}

We have found a family of stationary measures for the ASEP as seen
from a second class particle, parametrized by two numbers $\rho_-$ and
$\rho_+$, with $0\le\rho_-<\rho_+\le1$, which correspond to densities of the
asymptotic Bernoulli measures at $\pm\infty$.  The results, which hold for
all $p>1/2$, generalize those of \[djls], in which such measures were
obtained for the fully asymmetric case $p=1$.  Our derivation, in contrast to
that of previous works, is carried out directly in the infinite system.

{}For a certain value of the asymmetry, $x=x^*$, we found a two dimensional
representation of the algebra (see (\ref{xstar}) and (\ref{xstar1})).  In
fact, it can be shown that a $2r$ dimensional representation exists when
$x^r=x^*$, $r=1,2,\ldots$.  Finite dimensional representations of other
ASEP algebras have been found in \cite{Rit,MS}.

In the weak asymmetry limit $p\to1/2$, the measure can be understood in
terms of the solutions of the macroscopic viscous Burgers equation.
Several recent works have studied the gap of the evolution
operator  or generator~\[Kim,Nijs], as well as the diffusion constant~\[DK],
on a ring of size $L$, in the double limit $p\to1/2$, $L\to\infty$.  A simple
scaling form was found for the diffusion constant;  it would be nice to see
whether a scaling form in either problem could also be
derived from a macroscopic equation.

The microscopic shock profile derived in this work contains both intrinsic
features of the shock and the fluctuations of the position of the second
class particle. One can imagine many other ways of locating the
shock~\[inprep]; with an alternate definition, the expression for the
profile would certainly be different. However, one expects that there exist
intrinsic properties of the shock which are independent of the
definition of its location.  One class of such
properties would be the values of sums of expectations of all translates of
quantities whose expectation values vanish in the asymptotic Bernoulli
measures with densitites $\rho_+$ and $\rho_-$, e.g., 
 \begin{equation}
\sum_{i=-\infty}^\infty \langle (\tau_i - \rho_-) \tau_{i+n}
( \tau_{i+n+m} - \rho_+) \rangle. \label{Bsum}
 \end{equation}
 We have evaluated explicitly such sums  
for the totally asymmetric case $p=1$, using the invariant
measure as seen from the second class particle, and
will report on this work in a later publication~\[inprep].  
{}From this point of view, identities such as \(id1) reflect the fact that the
second class particle can be considered, in the sum, as either a first class
particle or a hole.

\ack
We are grateful to Errico Presutti for helpful discussions.  J.L.L. was
supported in part by AFOSR Grant 92--J--0115 and NSF Grant DMR 95-23266.
B.D. and J.L.L. thank DIMACS and its supporting agencies, the NSF under
contract STC-91-19999 and the N.J.  Commission on Science and Technology.
J.L.L. and E.R.S thank the Institut des Hautes Etudes Scientifiques and the
Ecole Normale Sup\'erieure for hospitality.

\appendix {Verification that the algebra yields an invariant measure}

In this appendix we show that if the operators $D$, $E$, and $A$ and the
vectors $\v$ and $\w$ satisfy the algebra (\ref{alg1})--(\ref{alg6}) then the
formula \(weight) yields an invariant measure for the ASEP as seen from a
second class particle.  We begin by writing down the conditions which must be
verified.  The algebra allows us to calculate directly the probability of
finding configurations  $\tau_{-m},\ldots,\tau_{-1}$ to the left, and
$\tau_1,\ldots,\tau_n$ to the right, of the second class particle, where the
$\tau_i$ are thought of as elements of the set $\{0,1\}$.
{}For convenience in describing the dynamics and consistency
with the notation of \(hop) we will write this local configuration as
$\tau=(\tau_{-m},\ldots,\tau_{-1},\tau_0,\tau_1,\ldots,\tau_n)$; in this
notation the symbol $\tau_0$ always takes the value $\tau_0=2$ and 
represents the presence of a second class particle at the origin.

Suppose then that for each such $\tau$ we have defined a weight $W(\tau)$. 
Then these weights define an invariant probability measure for the ASEP if
they satisfy (i)~normalization conditions:
 \begin{eqnarray}
   W(\tau_0) &=&
1,\nonumber\\ W(\tau_{-m},\ldots,\tau_{-1},\tau_0,\tau_1,\ldots,\tau_n) &=&
\sum_{\sigma=0,1}
W(\tau_{-m},\ldots,\tau_{-1},\tau_0,\tau_1,\ldots,\tau_n,\sigma )
    \label{norm}\\
 &=&  \sum_{\sigma=0,1}
W(\sigma,\tau_{-m},\ldots,\tau_{-1},\tau_0,\tau_1,\ldots,\tau_n);\nonumber
 \end{eqnarray}
 (ii)~positivity: 
 \begin{equation}
W(\tau_{-m},\ldots,\tau_{-1},\tau_0,\tau_1,\ldots,\tau_n)\ge0\;;
 \label{pos}
 \end{equation}
 and (iii)~for each configuration $\tau$ a condition corresponding to the
stationarity of the probability of $\tau$.  To write down these
stationarity conditions we will use the following notation: if
$\tau=(\tau_{-m},\ldots,\tau_n)$ is a system configuration and
$-m\le i\le n-1$ then $\tau^{i,i+1}$ denotes the configuration obtained from
$\tau$ by a particle jump across the bond between sites $i$ and $i+1$; if
$i,i+1\ne0$ this is an interchange of $\tau_i$ and $\tau_{i+1}$, but if either
$i=0$ or $i+1=0$ it is a shift of the configuration relative to the second
class particle:
 \begin{eqnarray}
\tau^{0,1}_j&=&\cases{\tau_1,&if $j=-1$,\cr
 2,&if $j=0$,\cr\tau_{j+1},&if $-m-1\le j\le-2$ or $1\le j\le n-1$;\cr}
   \nonumber\\
\tau^{-1,0}_j&=&\cases{\tau_{-1},&if $j=1$,\cr
 2,&if $j=0$,\cr\tau_{j-1},&if $-m+1\le j\le-1$ or $2\le j\le n+1$;\cr}\\
\tau^{i,i+1}_j&=&\cases{\tau_{i+1},&if $j=i$,\cr
 \tau_i,&if $j=i+1$,\cr\tau_j,&if $j\ne i,i+1$,\cr}\qquad 
   \hbox{when  $i,i+1\ne0$.}\nonumber
 \end{eqnarray}
   Now during some time
interval $dt$ there is, for each pair of adjacent sites $i,i+1$ with
$\tau_i\ne\tau_{i+1}$, some probability of exit from the configuration
$\tau$ by an interchange $\tau\to\tau^{i,i+1}$, and some probability
of entrance into $\tau$ by an interchange $\tau^{i,i+1}\to\tau$:
if $\tau_i\,\tau_{i+1}$ has the form $1\,0$, $1\,2$, or
$2\,0$ then the exit probability is $p\,dt$ and the entrance probability
$q\,dt$, while if $\tau_i\,\tau_{i+1}$ has the form $0\,1$, $2\,1$, or
$0\,2$ then these probabilities are reversed (see \(hop)).  
Moreover, there are
probabilities for exit from and entrance to $\tau$ due to exchanges across
the ends of $\tau$, between sites $-m$ and $-m-1$ or $n$ and $n+1$; if
$n=0$ or $m=0$, these exchanges involve the second class particle.
Combining these effects we find the equation
 \begin{eqnarray}
 0 &=&\sp_{i=-m}^{n-1} [-pW(\tau)+qW(\tau^{i,i+1})]
  +\spp_{i=-m}^{n-1}  [-qW(\tau)+pW(\tau^{i,i+1})]\nonumber\\
 &&\hskip20pt +\chi_1(\tau_{-m})[-qW(0\tau)+pW((0\tau)^{-m-1,-m})]\nonumber\\
 &&\hskip20pt +\chi_0(\tau_{-m})[-pW(1\tau)+qW((1\tau)^{-m-1,-m})]
    \label{bal}\\
 &&\hskip20pt +\chi_1(\tau_{n})  [-pW(\tau0)+qW((\tau0)^{n,n+1})]\nonumber\\
 &&\hskip20pt +\chi_0(\tau_{n})[-qW(\tau1)+pW((\tau1)^{n,n+1})]\;,\nonumber
  \end{eqnarray}
where the singly (respectively doubly) primed sum is over indices $i$ for
which $\tau_i\tau_{i+1}$ is $10$, $12$ or $20$ (respectively $01$, $21$ or
$02$), and the indicator functions $\chi_1$ and $\chi_0$ are defined by
 \begin{equation}
\chi_1(\sigma)
   =\cases{1,& if $\sigma=1$ or $\sigma=2$,\cr0,& if $\sigma=0$,\cr}\qquad
\chi_0(\sigma)=\cases{1,& if $\sigma=0$ or $\sigma=2$,\cr0,& if $\sigma=1$.\cr}
 \end{equation}

Equation \(bal) demands special attention if all the $\tau_i$ other than
$\tau_0$ are the same, say $\tau_i\equiv1$, $i\ne0$.  Then exchanges involving
the second class particle will not always change the local configuration.
{}For example, if $\tau=(1,1,1,2,1,1)$ then after the exchange
$\tau\to\tau^{0,1}=(1,1,1,1,2,1)$ the local configuration is still $\tau$
if there was originally a particle on site 3 (the third site to the right
of the origin). Thus \(bal), which includes a term $-qW(\tau)$ from this
exchange, should be corrected by a term $+qW(\tau1)$.  But \(bal) also
contains a term $+qW(\tau^{-1,0})$ from the exchange
$\tau^{-1,0}=(1,1,2,1,1,1)\to\tau$; in this case the local configuration
was already $\tau$ if there was before the exchange a particle at site
$-3$.  Thus \(bal) should also be corrected by a term
$-qW((1\tau)^{-1,0})$.  Since $\tau1=(1\tau)^{-1,0}$, these two corrections
cancel. Said otherwise, \(bal) counts the transition
$(1,1,1,2,1,1,1)\to(1,1,1,1,2,1,1)$, which does not change $\tau$, twice,
with opposite signs.  The argument is easily seen to be quite general and
to apply also to leftward jumps of the second class particle and to the
case $\tau_i\equiv0$, $i\ne0$.  We conclude that \(bal) is correct in all
cases.

We will now show that the weights defined by \(weight),
 \begin{equation}
   W_0(\tau_{-m},\ldots,\tau_{-1},\tau_0,\tau_1,\ldots,\tau_n)
 = \Ewv{\prod_{i=-m}^{-1}[\tau_i D+(1-\tau_i)E]\ A \ 
  \prod_{j=1}^n [\tau_j D+(1 -\tau_j)E]},\label{fund}
 \end{equation}
 satisfy (\ref{norm}), \(pos), and (\ref{bal}) and hence provide an invariant
measure for the ASEP.  (\ref{norm}) follows immedately from
(\ref{alg4})--(\ref{alg6}).  We will verify \(pos) by induction on $m+n$, the
total number of $D$ and $E$ operators in the product.  Since $\Ewv{A}=1$ by
\(alg6) the case $m+n=0$ is immediate.  For general $m+n$ it suffices to show
that
 \begin{equation}
 \Ewv{E^mAD^n}\ge0,  \label{easypos}
 \end{equation}
 since by repeated use of the relations (\ref{alg1})--(\ref{alg3}) we may
express the matrix element of any product with $m$ operators $E$ and $n$
operators $D$ as $x^k\Ewv{E^mAD^n}$, for some $k$, plus a linear
combination, with positive coefficients, of matrix elements of products with
a smaller number of $D$ or $E$ operators.  

To verify \(easypos) we define new operators $\d$ and $\e$ by
$D=\rho_+\rho_-+\d$ and\\ $E=(1-\rho_+)(1-\rho_-)+\e$ and show, again by
induction, that 
 \begin{equation}
 \Ewv{\e^mA\d^n}>0,  \label{easierpos}
 \end{equation}
 from which \(easypos) (with strict inequality) follows immediately.  The
operators $\d$ and $\e$ satisfy
\begin{eqnarray}
 \d\e - x\e\d =(1-x)ab,\ \ \
 \d A - xA\d =0,\ \ \        
 A\e - x\e A =0,&&\nonumber\\
 (\d+\e)\v = (a+b)\v,\ \ \ 
 \w(\d+\e)=\w(a+b), \hskip35pt\nonumber &&
 \end{eqnarray}
 and hence also $\d^k\e-x^k\e\d^k=ab(1-x^k)\d^{k-1}$ and
$\d\e^k-x^k\e^k\d=ab(1-x^k)\e^{k-1}$ for any $k\ge1$.  Then an argument as in
\(reduce) leads to the recursions
 \begin{eqnarray}
  \Ewv{\e^{m+1}A}&=&{1\over1-x^{m+2}}\Bigl[(a+b)(1-x^{m+1})\Ewv{\e^mA}
    \nonumber\\
   &&\hskip100pt-ab(1-x^m)\Ewv{\e^{m-1}A}\Bigr],\label{rec1}\\
 \Ewv{\e^mA\d^{n+1}}&=&(a+b)\Ewv{\e^mA\d^n}
  -x^{n+1}\Ewv{\e^{m+1}A\d^n}\nonumber\\
   &&\hskip100pt - ab(1-x^n)\Ewv{\e^mA\d^{n-1}};\label{rec2}
 \end{eqnarray}
 similar recursions hold with $m$ and $n$ interchanged, hence 
$\Ewv{\e^mA\d^n}= \Ewv{\e^nA\d^m}$. 
{}From \(rec1) it follows by induction on $m$ that
 \begin{equation}
  \Ewv{\e^mA} = {a^{m+1}-b^{m+1}\over a-b}{1-x\over1-x^{m+1}},\label{n0}
 \end{equation}
 and then from \(rec2) and \(n0), by induction on $n$, that for $n\le m$,
 \begin{equation}
 \Ewv{\e^mA\d^n}=\sum_{k=0}^n(ab)^k\ {a^{m+n+1-2k}-b^{m+n+1-2k}\over a-b}
  \  {(1-x)\prod_{j=k+1}^n(1-x^j) \over \prod_{j=0}^{n-k} (1-x^{m+1+j})}
  \label{final}
 \end{equation}
 (the identity $(a+b)(a^j-b^j)=(a^{j+1}-b^{j+1})+ab(a^{j-1}-b^{j-1})$ is
needed in checking \(n0) and \(final)).  The expression \(final) is clearly
positive, so that \(easierpos) holds.

Finally, we must verify (\ref{bal}).  To do so,
we partition the matrix product in \(fund)
into blocks of
consecutive identical matrices---that is, blocks of $D$'s or
$E$'s, together with one block consisting of a single $A$.
When the form (\ref{fund}) is substituted into the stationarity condition
(\ref{bal}), each resulting term arises from a possible exchange at some
block boundary---between blocks or at the end or beginning of the
product---and contains a corresponding factor
$\pm(pDE - qED)$,
$\pm(pAE - qEA)$, or
$\pm(pDA -qAD)$. These factors may be simplified with
the fundamental relations (\ref{alg1})--(\ref{alg3}), 
which we will use in several forms:
 \begin{eqnarray}
 p D E -q E D
 &=& (p-q) (eD + dE)\nonumber\\
 &=& (p-q) ((d-e)E + e(D+E))\nonumber\\
 &=& (p-q) ((e-d)D + d(D+E)), \\
 p A  E -q E A  &=& (p-q)eA,
      \nonumber\\
 p D A -q A D  &=& (p-q) dA. \nonumber      
\end{eqnarray}
 Here $d=\rho_+\rho_-$ and $e=(1- \rho_-)(1-\rho_+)$.

 The net effect of this simplification on each possible block boundary is
summarized in the following table:

 \vskip10pt
\ifdraft\openup-3\jot\fi
\def\lff{$\hfill\,$\cdots$\,\hfill$}
\tabskip0pt
\hbox to\hsize{\hss\vbox{%
 \halign{\tabskip30pt$#$
     &$\langle w|#|v\rangle$
     &\tabskip0pt\hss\hbox to 165pt
    {\hbox to 30pt{$(p-q)$}$\langle w|#|v\rangle$}\hss\cr
 \omit \hfill Boundary\hfill&\omit \hfill Contribution to (\ref{bal})\hfill&
       \omit\hfill Simplified contribution\hfill\cr
\noalign{\smallskip\hrule\smallskip}
 \hss DE\hss&
     \lff(-pDE+qED)\lff
     &\lff(-eD-dE)\lff\cr
 \hss ED\hss&
     \lff(+pDE-qED)\lff
     &\lff(+eD+dE)\lff\cr
 \hss DA\hss&
     \lff(-pDA+qAD)\lff
     &\lff(-dA)\lff\cr
 \hss AD\hss&
     \lff(+pDA-qAD)\lff
     &\lff(+dA)\lff\cr
 \hss AE\hss&
     \lff(-pAE+qEA)\lff
     &\lff(-eA)\lff\cr
 \hss EA\hss&
     \lff(+pAE-qEA)\lff
     &\lff(+eA)\lff\cr
 \hss\langle w|D\hss&
     (+pDE-qED)\lff
     &((e-d)D+d)\lff\cr
 \hss\langle w|E\hss&
     (-pDE+qED)\lff
     &((e-d)E-e)\lff\cr
 \hss\langle w|A\hss&
     (-pDA+qAD
     +pAE-qEA)\lff
     &((e-d)A)\lff\cr
 \hss D|v\rangle\hss&
     \lff(-pDE+qED)
     &\lff((d-e)D-d)\cr
 \hss E|v\rangle\hss&
     \lff(+pDE-qED)
     &\lff((d-e)E+e)\cr
 \hss A|v\rangle\hss&
     \lff(+pDA-qAD
     -pAE+qEA)
     &\lff((d-e)A)\cr
}}\hss}
\ifdraft\openup3\jot\fi

\vskip10pt
\noindent
The last column shows that each block of $D$'s gives rise to two
terms, one from each boundary of the block, in which one of the factors of
$D$ is replaced by the constant $d$; these terms have opposite signs
and hence cancel. Two similarly cancelling terms arise from each block of
$E$'s. Finally, the left and right ends of the product give rise to
additional terms in which the original amplitude is multiplied by $(e-d)$
and $(d-e)$, respectively; the cancellation of these terms completes the
verification of (\ref{bal}) for the weights (\ref{fund}).

\appendix {The invariant measure in the weakly asymmetric limit}

In this appendix we verify the picture given in Section~7 of the weakly
asymmetric limit of the invariant measure: for $\epsilon=p-q$ very small, the
measure is approximately a convolution of a Bernoulli measure having density
profile $\rho(y)$ with the density $Q(y)$ of the position of the second class
particle in this profile.  Here $y=\epsilon n$, so that $\rho(y)$ and $Q(y)$
(see \(tanh) and \(dens)) vary on the macroscopic scale . 

We begin by describing the key steps in the argument.  The probability of
occupation numbers $\sigma_1,\sigma_2,\ldots\sigma_m$ at specified sites
$n_1<n_2<\cdots<n_m$ is given by
 \begin{eqnarray}
\Ewv{(\sigma_1D+(1-\sigma_1)E)(D+E)^{n_2-n_1-1}
  (\sigma_2D+(1-\sigma_2)E)(D+E)^{n_3-n_2-1}\cdots 
   A\cdots &&\nonumber\\
  \cdots  (D+E)^{n_m-n_{m-1}-1}(\sigma_mD+(1-\sigma_m)E)}\hskip30pt
   &&\label{comp}
 \end{eqnarray}
 We want to compute an approximation to \(comp) that is correct in the
$\epsilon\to0$ limit.  First, each factor $(D+E)^n$ in the product has matrix
elements $(D+E)^n_{ij}=P_n(i-j)$; since $P_n(k)$ is the probability
distribution of a biased random walker and is concentrated near $k=n(a-b)$,
we approximate $(D+E)^{n-1}_{ij}\simeq(D+E)^n_{ij}\simeq\delta_{i-j,\lfloor
n(a-b)\rfloor}$.  Next, we approximate each single factor of $D$, $E$, or $A$
in the product by a diagonal matrix, obtained from equations \(delA),
\(epsA), and \(alphB) by making first the approximation $L\simeq L^{-1}\simeq
I$ and then a simple numerical approximation:
 \begin{eqnarray}
D_{ij} &\simeq& \left(\rho_+-2\lambda{x^j\over1+x^j}\right)\delta_{ij}
  \  \simeq\ \rho\left(\epsilon j\over2\lambda\right)\delta_{ij},
     \label{Dapprox}\\
E_{ij} &\simeq& \left(1-\rho_++2\lambda{x^j\over1+x^j}\right)\delta_{ij}
  \  \simeq\ \left(1- \rho\left(\epsilon j\over2\lambda\right)\right)
    \delta_{ij},
     \label{Eapprox}\\
A_{ij} &\simeq& -S_{jj}\delta_{ij}
  \  \simeq\ {\epsilon\over2\lambda}
   Q\left(\epsilon j\over2\lambda\right)\delta_{ij}.
     \label{Aapprox}
 \end{eqnarray}
With these approximations the inner product \(comp) becomes
 \begin{equation}
\sum_{j=-\infty}^\infty \prod_{k=1}^m
  \left[\sigma_k\rho\left(\epsilon (j+n_k)\over2\lambda\right) + 
 (1-\sigma_k)
  \left(1-\rho\left(\epsilon(j+n_k)\over2\lambda\right)\right)\right] 
  Q\left(\epsilon j\over2\lambda\right)\ {\epsilon\over2\lambda}\ .
 \label{Rsum}
 \end{equation}
 But this is just a Riemann sum for the integral
 \begin{equation}
\int_{-\infty}^\infty Q(z)\,dz\;\prod_{k=1}^m
   \left[\sigma_k\rho(y_k+z)+ (1-\sigma_k)(1-\rho(y_k+z)\right]\ ,
  \label{int}
 \end{equation}
  where $y_k=\epsilon n_k$.  Since the integrand
vanishes exponentially fast at $\pm\infty$ there is no difficulty with
convergence of the Riemann sums.  Moreover, we will sketch below a proof that
the sum of the errors made in the approximations leading to \(Rsum)---that
is, the difference between \(comp) and \(Rsum)---vanishes as
$\epsilon\to0$, uniformly in the choices of $n_k$ and $\sigma_k$. 
Thus we conclude: if the $n_k$ are chosen to depend on $\epsilon$ in such a
way that $\lim_{\epsilon\to0}\epsilon n_k=y_k$ then
 \begin{eqnarray} 
 &&\lim_{\epsilon\to0}\Pr\bigl\{\tau_{n_k}=\sigma_k\mid k=1,\ldots,m\bigr\}
  \nonumber\\
  &&\hskip80pt=\ \int_{-\infty}^\infty Q(z)\,dz\;\prod_{k=1}^m
   \left[\sigma_k\rho(y_k+z)+ (1-\sigma_k)(1-\rho(y_k+z)\right]\ .
 \label{main}
 \end{eqnarray}

 Equation \(main) gives the $\epsilon\to0$
limiting behavior of the profile, equation \(one), and the similar behavior
\(7sites) in the example discussed in Section~7 for the distribution of
the occupation numbers of several sites.  In general, \(main) describes the
limiting behavior of the distribution of occupation numbers at a finite
number of sites with both microscopic and macroscopic spacing, since we do
not assume that the positions $y_k$ are distinct. 

We now discuss the coarse grained random variables $B_\epsilon(y)$, the
empirical densities on an $\epsilon^{-\beta}$ scale, where $0<\beta<1$ (see
\(Bdef)).  We want to show that for a typical configuration in the
$\epsilon\to0$ limit these variables lie on the hyperbolic tangent profile
$\rho(y)$, after a configuration-dependent translation.  Due to this random
translation, however, the $B_\epsilon$ variables fluctuate even in the
$\epsilon\to0$ limit; to obtain a sharp statement, we show that there are no
fluctuations in a function of several of these variables,
$g(B)=g(B_\epsilon(y_1),\ldots,B_\epsilon(y_K))$, when $g$ is independent of
translations of the $y$ variables (see \(ti) below).  One example of this
technique is \(corr).  To show the absence of fluctuations in $g(B)$ we want
to show that $\E{|g(B)-\E{g(B)}|}=0$; the idea then is to use \(main) to
compute expectations of functions of the $B_\epsilon(y)$. 

Specifically, suppose that $g(\xi_1,\ldots,\xi_K)$ is a function defined for
$0\le\xi_k\le1$ and that $y_1,\ldots y_K$ are any real numbers.  We will show
that 
 \begin{equation}
\lim_{\epsilon\to0}\E{g(B_\epsilon(y_1),\ldots,B_\epsilon(y_K))}
  = \int_{-\infty}^{\infty}g(\rho(y_1+z),\ldots,\rho(y_K+z))\;Q(z)\,dz.
 \label{bins}
 \end{equation}
 We verify \(bins) first for $g$ a monomial, then a polynomial, then
continuous, and finally for $g$ bounded and continuous on
$(\rho_-,\rho_+)^K$, the case needed in applications.  If $g$ is a monomial,
say of degree $N$, then we may substitute the definition \(Bdef) of
$B_\epsilon$ into $g$ and expand, so that
$\E{g(B_\epsilon(y_1),\ldots,B_\epsilon(y_K))}$ becomes a sum of
$\bigl\lfloor\epsilon^{-\beta}\bigr\rfloor^N$ terms
$\epsilon^{-N\beta}\E{\prod_{j=1}^N\tau_{m_j}}$.  There are at most
${N\choose 2}\epsilon^{-(N-1)\beta}$ such terms in which not all the
$\tau_{m_j}$ are distinct, so that their contribution can be ignored, and
from \(main) the remaining terms give precisely the right hand side of
\(bins).  Thus \(bins) holds if $g$ is a polynomial and hence, by a uniform
approximation argument, if $g$ is continuous on $[0,1]^K$.  If we apply the
latter result, for $K=1$, to a continuous function $h_\delta(\xi)$ satisfying
$0\le h_\delta(\xi)\le1$, $h_\delta(\xi)=1$ for
$\rho_-+2\delta\le\xi\le\rho_+-2\delta$, and $h_\delta(\xi)=0$ for
$\xi\le\rho_-+\delta$ and $\xi\ge\rho_+-\delta$ we have
 \begin{equation}
  \Pr\{\rho_-+\delta\le B_\epsilon(y)\le\rho_+-\delta\}
   \ge\E{h_\delta(B_\epsilon(y))}
   \xxx\int_{-\infty}^\infty h_\delta(\rho(y+z))\;Q(z)\,dz,
 \end{equation}
 so that $\Pr\{\rho_-+\delta\le B_\epsilon(y)\le\rho_+-\delta\}\to1$ as
$\delta,\epsilon\to0$.  Using this result, we may extend \(bins) to any
function $g$ which is bounded and is continuous on $(\rho_-,\rho_+)^K$.  To do
so, we restrict $g$ to $(\rho_-+\delta,\rho_+-\delta)^K$ and then apply
\(bins) to a continuous extension of this restriction satisfying the same
bound as $g$, obtaining \(bins) for $g$ with error which vanishes as
$\delta\to0$. 
 
Now consider a function $g$ as above, bounded and continuous on
$(\rho_-,\rho_+)^K$, such that for some $y_1,\ldots,y_K$,
$g(\rho(y_1),\ldots,\rho(y_K))$ is translation invariant: for any $z$,
 \begin{equation}
  g(\rho(y_1),\ldots,\rho(y_K))=g(\rho(y_1+z),\ldots,\rho(y_K+z)).\label{ti}
 \end{equation}
  Then 
 \begin{eqnarray}
 &&\hskip-20pt \lim_{\epsilon\to0}
  \EE{\;\bigl|g(B_\epsilon(y_1),\ldots,B_\epsilon(y_K))
  - g(\rho(y_1),\ldots,\rho(y_K))\;\bigr|}\nonumber\\
 &&\hskip10pt = \int_{-\infty}^{\infty}\bigl|g(\rho(y_1+z),\ldots,\rho(y_K+z))
  -g(\rho(y_1),\ldots,\rho(y_K))\bigr|\;Q(z)\,dz\nonumber\\
 &&\hskip10pt = 0\,,
 \label{inell1}
 \end{eqnarray}
 so that $\lim_{\epsilon\to0}g(B_\epsilon(y_1),\ldots,B_\epsilon(y_K))
  = g(\rho(y_1),\ldots,\rho(y_K))$ in probability.

{}From \(inell1) we obtain the result described at the end of Section~7: if for
any $y$ we define $g(\xi_1,\xi_2) = \xi_2-\rho(y+\rho^{-1}(\xi_1))$ and then
apply \(inell1) to $g(B_\epsilon(0),B_\epsilon(y))$, we obtain \(eta).  Note
that $g$ is bounded but is not continuous at $\xi_1=\xi_2=\rho_-$ or
$\xi_1=\xi_2=\rho_+$. 

We finally sketch the argument that the errors made in the approximations
leading to \(Rsum) are uniformly small.  
Control of errors introduced by the approximation
 $(D+E)^n_{ij}\simeq\delta_{i-j,\lfloor n(a-b)\rfloor}$ is straightforward,
and details will be omitted. 
To  discuss
\(Dapprox)--\(Aapprox), we  simplify the notation by supposing that $n_1>0$ . 
Let us write $D^{(0)}$ for the diagonal approximation to $D$ introduced in
\(Dapprox), and $D^{(u)}=uD+(1-u)D^{(0)}$ for $0\le u\le1$, with similar
notation for $E$ and $A$.  We must then estimate
 \begin{eqnarray}
 && \int_0^1{d\over du}\Ewv{A^{(u)}(D+E)^{n_1-1}
   (\sigma_1D^{(u)}+(1-\sigma_1)E^{(u)})\ \cdots\ }\;du\nonumber\\
 &&\hskip50pt 
 = \sum_{j=1}^m \int_0^1\Ewv{A^{(u)}\ \cdots\ 
   (\sigma_j[D-D^{(0)}]+(1-\sigma_j)[E-E^{(0)}])\ \cdots\ }\;du\nonumber\\
 &&\hskip90pt 
  + \int_0^1\Ewv{[A-A^{(0)}](D+E)^{n_1-1}\ 
    \cdots\ }\;du\ .\label{sum2}
 \end{eqnarray}
 On the right hand side of \(sum2) we regard $\w$ and $\v$ as elements (of
norm~1) of $\ell^\infty=\ell^\infty(\bbz)$, and note that $D$, $E$,
$D^{(0)}$, $E^{(0)}$, and $D+E$ are bounded operators on $\ell^\infty$, of
norm at most $1$, and that $A$ and $A^{(0)}$ are bounded operators from
$\ell^\infty$ to $\ell^1(\bbz)$.  Now consider one term from the sum over $j$
in \(sum2), for which the matrix product will contain a factor of either
$D-D^{(0)}$ or $E-E^{(0)}$.  Each of these may be written as the sum of two
terms, corresponding to the two approximations made in (\ref{Dapprox}) or
(\ref{Eapprox}).  The first of these terms contains a factor $L-I$, which we
carry to the right until it reaches, and annhilates, $\v$; $L-I$ commutes
with $D+E$ and its commutator with $D^{(u)}$ or $E^{(u)}$ has $\ell^\infty$
operator norm which is $O(\epsilon)$.  The second error term also has
$\ell^\infty$ operator norm which is $O(\epsilon)$.  The contribution to
\(sum2) containing $A-A^{(u)}$ is treated similarly, and we derive an overall
bound for \(sum2) which is a constant multiple of $m^2\epsilon$.

\clearpage

\centerline{FIGURE CAPTIONS}
\bigskip
\bigskip
\noindent
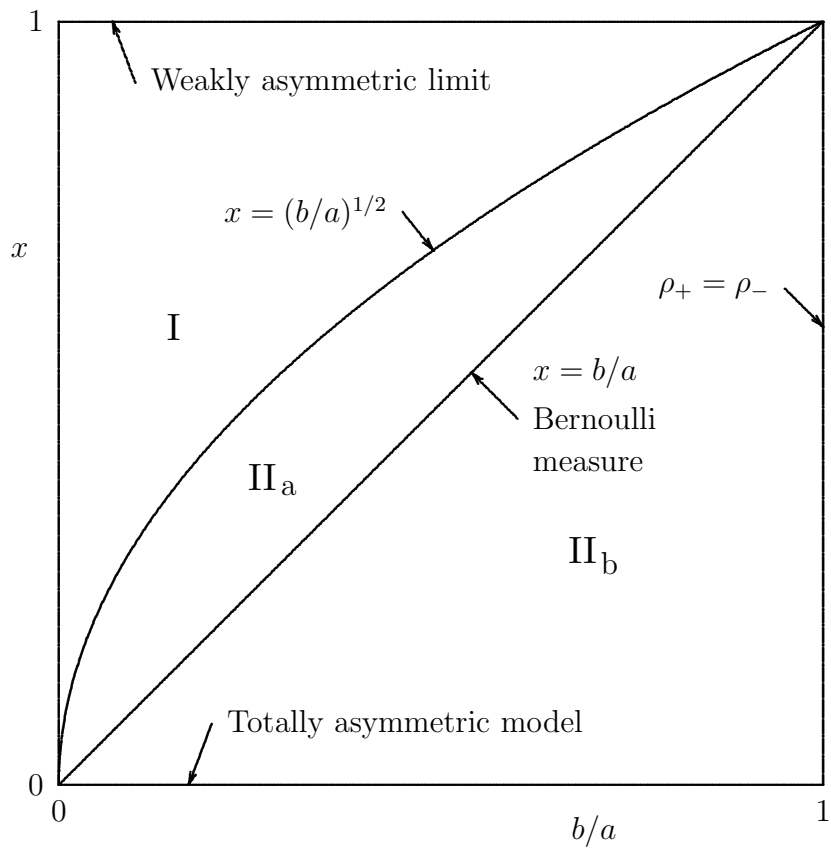
{\bf Figure 1.} A phase diagram for the asymptotics of the shock in the
ASEP.  For $n>>1$,
$\langle\tau_n\rangle\simeq \rho_++Cn^{\gamma}\exp-\kappa n$, with
$\kappa=-\log(1-a-b+ax+b/x)$ and $\gamma=1$ in region I and
$\kappa=-\log(1-(\sqrt a-\sqrt b)^2)$ and $\gamma=-3/2$ in region II, and
with $C<0$ in regions I and $\rm II_a$ and $C>0$ in region $\rm II_b$.

 \medskip
 \noindent
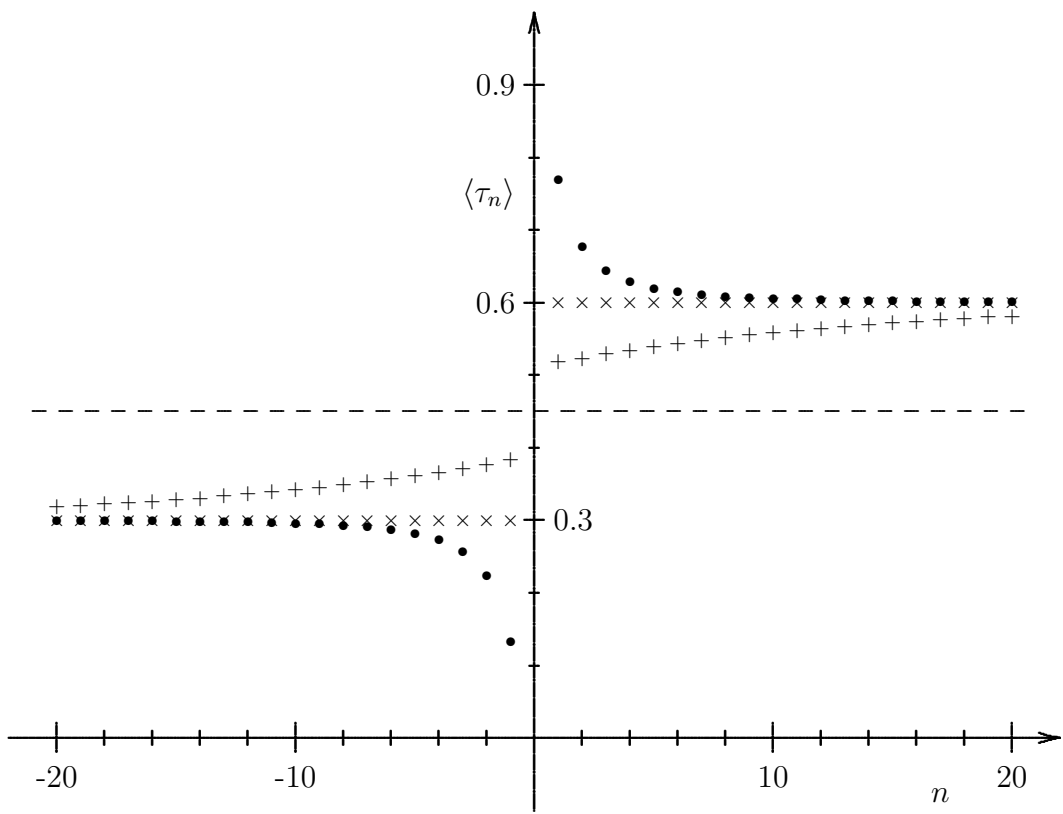
 {\bf Figure 2.} Profile $\langle\tau_n\rangle$ for $\rho_-=0.3$,
$\rho_+=0.6$: $x=0.0$ ($\bullet$); $x=x^*=0.2857$ ($\times$); $x=0.6$
($+$).  The dashed line is at height $(\rho_-+\rho_+)/2$.

%% BELOW IS TEX CODE FOR THE FIGURES.  IT REQUIRES THE MACRO PACKAGE
%% pictex.tex .   IF THIS PACKAGE IS NOT AVAILABLE OR IF THE FIGURES ARE
%% NOT WANTED, SIMPLY UNCOMMENT THE NEXT LINE TO END THE DOCUMENT AT THIS
%% POINT. 

%% \end{document}

\clearpage

\thispagestyle{empty}
\let\fiverm\relax
\input pictex

%% FIGURE 1

\null\vfill

 \font\bigfont cmr10 at 14.4truept
 \setplotsymbol({$\scriptstyle.$})
 \centerline{\beginpicture
 \setcoordinatesystem units <4truein,4truein> point at 0 0
 \setplotarea x from 0.0 to 1.1, y from 0.0 to 1.1
 \setlinear
  \plot 0.0 1.0 1.0 1.0 1.0 0.0 0.0 0.0 0.0 1.0 /
 \plot 0.0 0.0 1.0 1.0 /
 \put {0} [t] at 0 -0.02
 \put {1} [t] at 1 -0.02
 \put {0} [r] at -0.02 0
 \put {1} [r] at -0.02 1
 \put {$b/a$}  [t] at 0.7 -0.04 
 \put {$x$}  [r] at -0.04 0.7
 \put {Totally asymmetric model} [l] at 0.22 0.08
 \arrow <5pt> [0.2,0.5] from 0.20 0.08 to 0.17 0.00
 \put {$\rho_+=\rho_-$} [r] at 0.93 0.65
 \arrow <5pt> [0.2,0.5] from 0.95 0.65 to 1.0 0.6
 \put {Weakly  asymmetric limit} [l] at 0.12 0.92
 \arrow <5pt> [0.2,0.5] from 0.10 0.92 to 0.07 1.00
 \put {$x=b/a$} [l] at 0.62 0.54
 \put {Bernoulli} [l] at 0.62 0.48
 \put {measure} [l] at 0.62 0.42
 \arrow <5pt> [0.2,0.5] from 0.6 0.48 to 0.54 0.54
 \put {$x=(b/a)^{1/2}$} [r] at 0.43 0.75
 \arrow <5pt> [0.2,0.5] from 0.45 0.75 to 0.49 0.70
 \put {$\hbox{\bigfont II}_{\,\hbox{b}}$} at  0.7 0.3
 \put {$\hbox{\bigfont II}_{\,\hbox{a}}$} at  0.28 0.4
 \put {\bigfont I} at  0.15 0.6
 \setquadratic 
 \plot  0.0 0.0 0.25 0.5 1.0 1.0 /
\endpicture}
\vfill
\hbox to\hsize{\hss Figure 1}
\clearpage

%% FIGURE 2

\thispagestyle{empty}
\null\vfill

\setplotsymbol({$\scriptstyle .$})

\centerline{
 \beginpicture
 \setcoordinatesystem units <0.125truein,3.8truein> point at 0 0
 \setplotarea x from -21.0 to 21.0, y from -0.1 to 1.1
%% Axes and labels
 \arrow <8pt>  [0.2, 0.5]  from -22.0 0.0  to 22.0 0.0 
 \arrow <8pt>  [0.2, 0.5]  from 0.0 -0.1  to 0.0 1.0
 \put {$n$} [t] at 17 -0.07 
 \put {$\langle\tau_n\rangle$} [r] at -0.8 0.75
%% x-axis tics and tic labels
 \plot 2 -0.01 2 0.01 /
 \plot 4 -0.01 4 0.01 /
 \plot 6 -0.01 6 0.01 /
 \plot 8 -0.01 8 0.01 /
 \plot 12 -0.01 12 0.01 /
 \plot 14 -0.01 14 0.01 /
 \plot 16 -0.01 16 0.01 /
 \plot 18 -0.01 18 0.01 /
 \plot -2 -0.01 -2 0.01 /
 \plot -4 -0.01 -4 0.01 /
 \plot -6 -0.01 -6 0.01 /
 \plot -8 -0.01 -8 0.01 /
 \plot -12 -0.01 -12 0.01 /
 \plot -14 -0.01 -14 0.01 /
 \plot -16 -0.01 -16 0.01 /
 \plot -18 -0.01 -18 0.01 /
 \plot 20 -0.02  20 0.02 /
 \put {20} [t] at 20 -0.04
 \plot -20 -0.02  -20 0.02 /
 \put {-20} [t] at -20 -0.04
 \plot 10 -0.02  10 0.02 /
 \put {10} [t] at 10 -0.04
 \plot -10 -0.02  -10 0.02 /
 \put {-10} [t] at -10 -0.04
%% y-axis tics and tic labels
 \plot -0.2 0.1 0.2 0.1 /
 \plot -0.2 0.2 0.2 0.2 /
 \plot -0.2 0.4 0.2 0.4 /
 \plot -0.2 0.5 0.2 0.5 /
 \plot -0.2 0.7 0.2 0.7 /
 \plot -0.2 0.8 0.2 0.8 /
 \plot -0.4 0.6 0.4 0.6 /
 \put {0.6} [r] at -0.8 0.6
 \plot -0.4 0.9 0.4 0.9 /
 \put {0.9} [r] at -0.8 0.9
 \plot -0.4 0.3 0.4 0.3 /
 \put {0.3} [l] at 0.8 0.3
%% Profiles
 \put {$\scriptstyle \bullet$} at    1   0.768
 \put {$\scriptstyle \bullet$} at    2   0.677
 \put {$\scriptstyle \bullet$} at    3   0.644
 \put {$\scriptstyle \bullet$} at    4   0.628
 \put {$\scriptstyle \bullet$} at    5   0.619
 \put {$\scriptstyle \bullet$} at    6   0.614
 \put {$\scriptstyle \bullet$} at    7   0.610
 \put {$\scriptstyle \bullet$} at    8   0.608
 \put {$\scriptstyle \bullet$} at    9   0.606
 \put {$\scriptstyle \bullet$} at   10   0.605
 \put {$\scriptstyle \bullet$} at   11   0.604
 \put {$\scriptstyle \bullet$} at   12   0.603
 \put {$\scriptstyle \bullet$} at   13   0.602
 \put {$\scriptstyle \bullet$} at   14   0.602
 \put {$\scriptstyle \bullet$} at   15   0.602
 \put {$\scriptstyle \bullet$} at   16   0.601
 \put {$\scriptstyle \bullet$} at   17   0.601
 \put {$\scriptstyle \bullet$} at   18   0.601
 \put {$\scriptstyle \bullet$} at   19   0.601
 \put {$\scriptstyle \bullet$} at   20   0.601
 \put {$\scriptstyle \bullet$} at   -1   0.132
 \put {$\scriptstyle \bullet$} at   -2   0.223
 \put {$\scriptstyle \bullet$} at   -3   0.256
 \put {$\scriptstyle \bullet$} at   -4   0.272
 \put {$\scriptstyle \bullet$} at   -5   0.281
 \put {$\scriptstyle \bullet$} at   -6   0.286
 \put {$\scriptstyle \bullet$} at   -7   0.290
 \put {$\scriptstyle \bullet$} at   -8   0.292
 \put {$\scriptstyle \bullet$} at   -9   0.294
 \put {$\scriptstyle \bullet$} at  -10   0.295
 \put {$\scriptstyle \bullet$} at  -11   0.296
 \put {$\scriptstyle \bullet$} at  -12   0.297
 \put {$\scriptstyle \bullet$} at  -13   0.298
 \put {$\scriptstyle \bullet$} at  -14   0.298
 \put {$\scriptstyle \bullet$} at  -15   0.298
 \put {$\scriptstyle \bullet$} at  -16   0.299
 \put {$\scriptstyle \bullet$} at  -17   0.299
 \put {$\scriptstyle \bullet$} at  -18   0.299
 \put {$\scriptstyle \bullet$} at  -19   0.299
 \put {$\scriptstyle \bullet$} at  -20   0.299
 \put {$\scriptstyle \times$} at    1   0.600
 \put {$\scriptstyle \times$} at    2   0.600
 \put {$\scriptstyle \times$} at    3   0.600
 \put {$\scriptstyle \times$} at    4   0.600
 \put {$\scriptstyle \times$} at    5   0.600
 \put {$\scriptstyle \times$} at    6   0.600
 \put {$\scriptstyle \times$} at    7   0.600
 \put {$\scriptstyle \times$} at    8   0.600
 \put {$\scriptstyle \times$} at    9   0.600
 \put {$\scriptstyle \times$} at   10   0.600
 \put {$\scriptstyle \times$} at   11   0.600
 \put {$\scriptstyle \times$} at   12   0.600
 \put {$\scriptstyle \times$} at   13   0.600
 \put {$\scriptstyle \times$} at   14   0.600
 \put {$\scriptstyle \times$} at   15   0.600
 \put {$\scriptstyle \times$} at   16   0.600
 \put {$\scriptstyle \times$} at   17   0.600
 \put {$\scriptstyle \times$} at   18   0.600
 \put {$\scriptstyle \times$} at   19   0.600
 \put {$\scriptstyle \times$} at   20   0.600
 \put {$\scriptstyle \times$} at   -1   0.300
 \put {$\scriptstyle \times$} at   -2   0.300
 \put {$\scriptstyle \times$} at   -3   0.300
 \put {$\scriptstyle \times$} at   -4   0.300
 \put {$\scriptstyle \times$} at   -5   0.300
 \put {$\scriptstyle \times$} at   -6   0.300
 \put {$\scriptstyle \times$} at   -7   0.300
 \put {$\scriptstyle \times$} at   -8   0.300
 \put {$\scriptstyle \times$} at   -9   0.300
 \put {$\scriptstyle \times$} at  -10   0.300
 \put {$\scriptstyle \times$} at  -11   0.300
 \put {$\scriptstyle \times$} at  -12   0.300
 \put {$\scriptstyle \times$} at  -13   0.300
 \put {$\scriptstyle \times$} at  -14   0.300
 \put {$\scriptstyle \times$} at  -15   0.300
 \put {$\scriptstyle \times$} at  -16   0.300
 \put {$\scriptstyle \times$} at  -17   0.300
 \put {$\scriptstyle \times$} at  -18   0.300
 \put {$\scriptstyle \times$} at  -19   0.300
 \put {$\scriptstyle \times$} at  -20   0.300
 \put {$\scriptstyle +$} at    1   0.518
 \put {$\scriptstyle +$} at    2   0.523
 \put {$\scriptstyle +$} at    3   0.529
 \put {$\scriptstyle +$} at    4   0.534
 \put {$\scriptstyle +$} at    5   0.539
 \put {$\scriptstyle +$} at    6   0.543
 \put {$\scriptstyle +$} at    7   0.547
 \put {$\scriptstyle +$} at    8   0.551
 \put {$\scriptstyle +$} at    9   0.555
 \put {$\scriptstyle +$} at   10   0.558
 \put {$\scriptstyle +$} at   11   0.561
 \put {$\scriptstyle +$} at   12   0.564
 \put {$\scriptstyle +$} at   13   0.567
 \put {$\scriptstyle +$} at   14   0.570
 \put {$\scriptstyle +$} at   15   0.572
 \put {$\scriptstyle +$} at   16   0.574
 \put {$\scriptstyle +$} at   17   0.576
 \put {$\scriptstyle +$} at   18   0.578
 \put {$\scriptstyle +$} at   19   0.580
 \put {$\scriptstyle +$} at   20   0.581
 \put {$\scriptstyle +$} at   -1   0.383
 \put {$\scriptstyle +$} at   -2   0.377
 \put {$\scriptstyle +$} at   -3   0.371
 \put {$\scriptstyle +$} at   -4   0.366
 \put {$\scriptstyle +$} at   -5   0.361
 \put {$\scriptstyle +$} at   -6   0.357
 \put {$\scriptstyle +$} at   -7   0.353
 \put {$\scriptstyle +$} at   -8   0.349
 \put {$\scriptstyle +$} at   -9   0.345
 \put {$\scriptstyle +$} at  -10   0.342
 \put {$\scriptstyle +$} at  -11   0.339
 \put {$\scriptstyle +$} at  -12   0.336
 \put {$\scriptstyle +$} at  -13   0.333
 \put {$\scriptstyle +$} at  -14   0.330
 \put {$\scriptstyle +$} at  -15   0.328
 \put {$\scriptstyle +$} at  -16   0.326
 \put {$\scriptstyle +$} at  -17   0.324
 \put {$\scriptstyle +$} at  -18   0.322
 \put {$\scriptstyle +$} at  -19   0.320
 \put {$\scriptstyle +$} at  -20   0.319
 \setdashes <5pt>
 \setplotsymbol({$\scriptscriptstyle .$})
 \plot -21 0.45 21 0.45 /
 \endpicture
}

\vfill
\hbox to\hsize{\hss Figure 2}

 \end{document}